\documentclass{elsarticle}
\usepackage{hyperref}
\usepackage[utf8]{inputenc}
\usepackage{graphicx}
\usepackage{textcomp}
\usepackage{amsmath}	% Advanced maths commands
\usepackage{amssymb}	% Extra maths symbols
\usepackage{lscape, longtable}	% Landscape pages
\journal{New Astronomy}

\biboptions{authoryear}

\begin{document}

\begin{frontmatter}

\title{Monitoring of transiting exoplanets and their host stars with small aperture telescopes}

\author{M.A.Salisbury, U.C.Kolb, A.J.Norton, C.A.Haswell}
\address{School of Physcial Sciences, The Open University, Milton Keynes, MK7 6AA, UK}


%\author[mysecondaryaddress]\corref{mycorrespondingauthor}
%\cortext[mycorrespondingauthor]{Mark Salisbury}
%\ead{mark.salisbury@open.ac.uk}


\begin{abstract}
Exoplanet research is now target rich with a wide diversity of systems making it difficult for high demand observatories to undertake follow up observations over extended periods of time.  We investigate the effectiveness of using 0.4m-class telescopes for monitoring transiting hot Jupiters and their host stars.  We consider two representative case studies:  WASP-52b with 13 new transits, and HAT-P-23b with 17 new transits and  concurrent photometric monitoring covering 78 days. We present updated system parameters and combine our new transit times  with previously published results to calculate new ephemerides for both systems.  Our analysis of transit mid-times for WASP-52b results in a slight preference for a quadratic ephemeris ($\Delta\chi_{\nu}^2 = 0.07$, $\Delta BIC = 1.53$ over a linear ephemeris.  We discuss the reality of this quadratic ephemeris indicating a period change of $ \delta P / \delta t = -38.6\pm4{\rm ms~yr^{-1}}$ and consider possible causes.  WASP-52 is known to be an active star with previous publications reporting many spot crossing events, however no such events are seen in our new photometry.  Our analysis shows that WASP-52 is still active and that the latitude of the spots has likely migrated away from the transit chord.  We confirm the inflated nature and circular orbit for HAT-P-23b.  Our monitoring of HAT-P-23 reveals a periodicity of 7.015 days with an amplitude of 0.011 mag which we interpret as the rotation period of HAT-P-23.  The photometric and transit timing precision achieved in the case studies shows that this class of telescope is capable of precise characterisation and long-term monitoring of transiting hot Jupiters in support of dedicated ongoing and future ground and space based observations.
\end{abstract}

\begin{keyword}
%\texttt{elsarticle.cls}\sep \LaTeX\sep Elsevier \sep template
%\MSC[2010] 00-01\sep  99-00
\end{keyword}

\end{frontmatter}

%\linenumbers

\section{Introduction}
We now know of 465 hot Jupiters\footnote{NASA Exoplanet Archive, accessed 3rd January 2020.}, defined as planets with $0.5R_{\rm{J}} < R_{\rm{P}} < 2R_{\rm{J}}$ and periods \textbf{$<$} 10d, and further discoveries are expected from SuperWASP, HATNet, NGTS, TESS and PLATO.  Exoplanet research is now target rich with a wide diversity of systems making it difficult for large, high demand, observatories to undertake follow up observations over extended periods of time.  Conversely small telescopes, including those associated with educational institutions and amateur operated observatories are  plentiful, for example there are 1125 listed contributors to the Exoplanet Transit Database maintained by the the Czech Astronomical Society.\footnote{http://var2.astro.cz/ETD/, accessed $\rm{11^{th}}$ February 2020}  These small observatories are available for the monitoring of bright transiting hot Jupiters and their host stars over extended time periods.  Such observations can provide invaluable insight into exoplanet system properties including the existence of companion bodies \citep[e.g.][]{Sokov2018TransitC} and the formation and dynamical histories of the systems due to orbital period evolution \citep{Maciejewski2016,Patra2017,Bouma2019WASP-4bMission,Chontos2019TheExoplanet}.  Here we illustrate the capabilities of small ($\sim$ 0.4meter) aperture telescopes to contribute to our understanding of exoplanet systems with our study of two known transiting exoplanets.  HAT-P-23b and WASP-52b were observed from June to October 2018 with the  OpenScience Observatories (OSO) facility.  The OSO are part of the Open University OpenSTEM Labs\footnote{learn5.open.ac.uk}used for teaching practical science to undergraduate distance learners and for postgraduate research.   Additional observations were made using a small private observatory (POST) located in the United Kingdom. 

\subsection{HAT-P-23b}
HAT-P-23\,b is a short period (1.21d) hot Jupiter discovered in 2011 by the HATNet transit survey orbiting a V=12.43 magnitude G0 host star.  Discovered by \cite{Bakos2011HAT-P-20b-HAT-P-23b:Planets}, they found HAT-P-23\,b is an inflated hot Jupiter with a mass of $2\rm{M_J}$ and a radius of $1.37\rm{R_J}$, almost 10\% greater than the maximum theoretical radius for a planet of its mass \cite{Fortney2007PlanetaryTransits}.  In addition they found HAT-P-23\,b has a marginally significant orbital eccentricity of $0.106\pm0.044$.  \cite{Ciceri2015} subsequently found considerably smaller stellar and planetary radii (Figure \ref{fig:HAT-P-23b_Exofast_Outputsv2.png}), which we can speculate results from adopting a mass-radius relationship based on a younger age for HAT-P-23 and using a fixed circular orbit in their transit modelling.  Rossiter-McLaughlin effects have shown HAT-P-23\,b is in an aligned prograde orbit with a projected angle between the orbital plane and the stellar equatorial plane of 15\textdegree$\pm\,22$\textdegree \citep{Moutou2011Spin-orbitHAT-P-23b}.  Previous studies have taken different approaches to the suspected eccentricity with some authors leaving eccentricity as a free parameter, e.g. \cite{Bakos2011HAT-P-20b-HAT-P-23b:Planets, Sada2016}, while others have fixed the orbit as circular in their analyses e.g. \cite{Ciceri2015, Maciejewski2018Planet-starB}.   HAT-P-23\,b has a similar orbital period to that of WASP-12\,b.  The latter has been shown to exhibit a decreasing orbit \citep{Maciejewski2016, Patra2017, Maciejewski2018Planet-starB, Yee2019TheDecaying} hence HAT-P-23\,b could be expected to exhibit a similar decreasing orbital period which may require transit measurements over decades to detect.  \cite{Ciceri2015} presented O-C results (their Figure 6) that appear to show a quadratic deviation indicative of period change, but this was not commented upon.  The transit timing observations available on the Exoplanet Transit Database\footnote{http://var2.astro.cz/ETD/} (ETD) show a change in reported O-C values over the period between transit epochs 1900 and 2050.  \cite{Maciejewski2018Planet-starB} published results of a further 13 new transits and reanalysis of the previously published transit times looking for evidence of orbital decay.  They conclude that the best fit ephemeris is linear and that there is no sign of orbital period change which they attributed to inefficient tidal dissipation in the main sequence host star. 

\subsection{WASP-52b}
WASP-52\,b is an inflated hot Jupiter on a 1.75d circular orbit around a K2V star \citep{Hebrard2013WASP-52bPlanets}.  The $0.79\rm{R}_{\odot}$ stellar radius and the inflated $1.27\rm{R_J}$ planet radius results in 3\% deep transits.  Due to its large radius and low mass the difference in transit depth of a single scale height equates to $4.4x10^{-4}$ \citep{Kirk2016TransmissionSurface}.  This makes WASP-52b an excellent target for transit spectroscopy; such studies reveal that WASP-52b has a cloudy atmosphere with no indication of Rayleigh scattering \citep{Kirk2016TransmissionSurface, Louden2017AWASP-52b}, in which sodium has been detected  \citep{Chen2017TheAtmosphere}.

WASP-52 is an active star with transit photometry often showing spot crossing events complicating the analysis of transit spectroscopy results. Photometrically determined rotational periods between $13.1\pm0.4\textrm{d}$ to $17.79\pm0.05\textrm{d}$ have been reported, possibly indicative of differential rotation at changing spot latitudes \citep{Hebrard2013WASP-52bPlanets, Mancini2017OrbitalSystem, Louden2017AWASP-52b}.  Measurements of crossings of the same spot has allowed the sky projected and real orbital obliquities to be determined as $3.8$\textdegree $\,\pm\,8.4$\textdegree and $20$\textdegree$\,\pm\,50$\textdegree  respectively, providing the first measurement of a true orbital obliquity using this method \citep{Mancini2017OrbitalSystem}. 

This paper is laid out as follows.  In Section 2 we describe the telescopes and methods used to obtain and analyse the observations.  Section 3 presents our analysis. Section 4 places our results in the context of the published literature. Section 5 discusses the performance of the small aperture telescopes and their capacity to contribute to the forefront of exoplanet research.  In Section 6 we summarise our findings.

\section{Methods}\label{Methods}

\subsection{OpenScience Observatories}
The OpenScience Observatories (OSO) consist of three robotic telescopes, two optical telescopes located at Observatorio del Teide, part of the Observatorios de Canarias on the island of Tenerife, 28\textdegree 18$^{\prime}$ 00$^{\prime\prime}$ N, 16\textdegree 30$^{\prime}$ 35$^{\prime\prime}$ W at an altitude of 2390m \citep{Kolb2018ATeaching}.  The third, ARROW a remotely controlled radio telescope based in Milton Keynes UK, was not used in this work.  The two telescopes in Tenerife are COAST, a 0.35m f11 Schmidt-Cassegrain optical system, and PIRATE, a 0.45m f6.8  Corrected Dall-Kirkham astrograph.  Each telescope is located in its own robotic dome on a German Equatorial mount which can be operated completely autonomously via a scheduler or remotely in real time.  Both observatories are controlled using the autonomous telescope control software, ABOT \citep{Sybilski2014SoftwareData}, the COAST telescope is also available for public use as part of telescope.org.  PIRATE is equipped with an FLI ProLine 16803 $4\rm{k} \times 4\rm{k}$  $9\mu$ pixel CCD camera producing a field of view $42^\prime$ square with plate scale of $0.62^{\prime \prime}$/pixel.  COAST is equiped with an FLI ProLine 9000 $3\rm{k} \times 3\rm{k}$ $12\mu$ pixel CCD camera producing a field of view $32^\prime$ square with plate scale of $0.63^{\prime\prime}$/pixel.  Both CCD cameras were operated at a temperature of -30\textcelsius.  In July 2018 both telescopes were upgraded with the addition of a GPS controlled shutter timing system.  All observations were made through Johnson B or Cousins Rc filters without guiding.  

\subsection{POST Observatory}
The POST observatory, located at 51\textdegree 15$^{\prime}$ 51$^{\prime\prime}$, 01\textdegree 14$^{\prime}$ 58$^{\prime\prime}$E at 20m altitude, hosts a 0.4m, f6.8 telescope on a German Equatorial mount.  It is equipped with an SBIG ST10-XME $ 2184 \times 1472$ $6.8\mu$ pixel CCD camera providing a field of view $19^\prime\times13^\prime$ with a plate scale of $0.52^{\prime\prime}$/pixel.  All observations were made using Johnson V or Cousins Rc filters.  Observations where the telescope was not defocussed were guided using an on-axis tip-tilt guiding corrector at rates between 0.1-2Hz depending on the brightness of the available guide star.  Where defocussed observations were made, off axis guiding was employed with correction rates $\sim$0.2Hz. 

\subsection{Observations}
Transit observations of both targets made using PIRATE and COAST were scheduled in advance and obtained in autonomous mode, observations with the POST observatory were made in attended mode with hands-on control of the telescope during each observation.  The observations were planned using existing ephemerides and timed to include pre and post transit data where possible.    Observations with all telescopes were made with the CCD on-chip binning of $2\times2$ or $1\times1$.  Exposure times were set to ensure ADU counts remained within the linear response region of the CCD for the target and comparison stars and the same exposure duration was used throughout each individual transit observation.  Full details of the observations made can be found in Table \ref{tab:Observations}.

Monitoring observations were obtained of HAT-P-23 using PIRATE scheduled to take up to six 150 second frames per night using the Rc filter with the CCD in 1x1 bin mode when the target was higher than 40 degrees altitude.  218 observations covering 39 nights over a 78 day period were made between $18^{{\rm th}}$ August 2018 and $4^{{\rm th}}$ November 2018.  The number of frames obtained each night varied depending on weather conditions and telescope schedule commitments.  Out-of-transit measurements from before and after each transit, where available, were added to the monitoring data to extend the measurement timeline and allow HAT-P-23's magnitude to be measured at the time of each transit recorded with PIRATE.

Dark, bias and sky flat-field calibration frames were obtained for each night of science observation.  Due to the limited number of dusk/dawn flat frames obtainable by PIRATE each night frames from multiple nights either side of the observation were combined. For POST a flat field light box is used so an average of 27 flat field frames was obtained on each night of observation.

As one of the key aims of the observations was to measure the transit mid-times a cross check was carried out between the timing accuracy of both the PIRATE and POST telescopes by simultaneously observing the deeply eclipsing post common envelope binary V1828Aql.  Unfiltered observations were made using the same 1 minute integration time on both telescopes.  The observations from both telescopes agreed on the eclipse minimum times to within 4 seconds for the two primary eclipses, well within typical Exoplanet transit mid-time uncertainties of 30 to 50 seconds. 

\subsection{Data Reduction and Photometry}
The calibration frames obtained for each night of science observation were median combined using the Data Processing module of AstroImageJ v3.2.1\citep{Collins2017}, hereafter AIJ, to create master calibration frames for each transit observation.  Individual calibration frames were visually inspected to ensure consistency and any with artifacts were discarded.  All science frames were plate solved and individual mid exposure times converted to BJD$_{\text{TDB}}$.

Ensemble photometry was carried out within AIJ using up to 25 comparison stars of suitable brightness from across the entire field of view.  The photometry was compared with a  transit model created in AIJ and the photometry apertures adjusted to minimise the residual RMS and Bayesian Information Criterion (BIC) values.  Where it produced superior results the sky aperture was varied by a fixed multiplier of the target stellar FWHM.  Following selection of the optimum aperture sizes each individual comparison star was removed from the ensemble in sequence to test if its inclusion increased or decreased the RMS and BIC values.  Stars found to increase these values were removed, typically leaving 15-20 comparison stars in the photometry ensemble for each transit data set.  Frames producing outlying data points were removed from the photometry where visual inspection revealed cosmetic effects such as cosmic ray strikes or satellite trails. 

Two archive observations, one of WASP-52\,b from 2013 and one of HAT-P-23\,b from 2017,  made with POST were also included in the analysis.  Both transits were re-analysed from the original images and calibration files in the same way as the newly obtained data.

\begin{figure}
 \includegraphics[width=\textwidth]{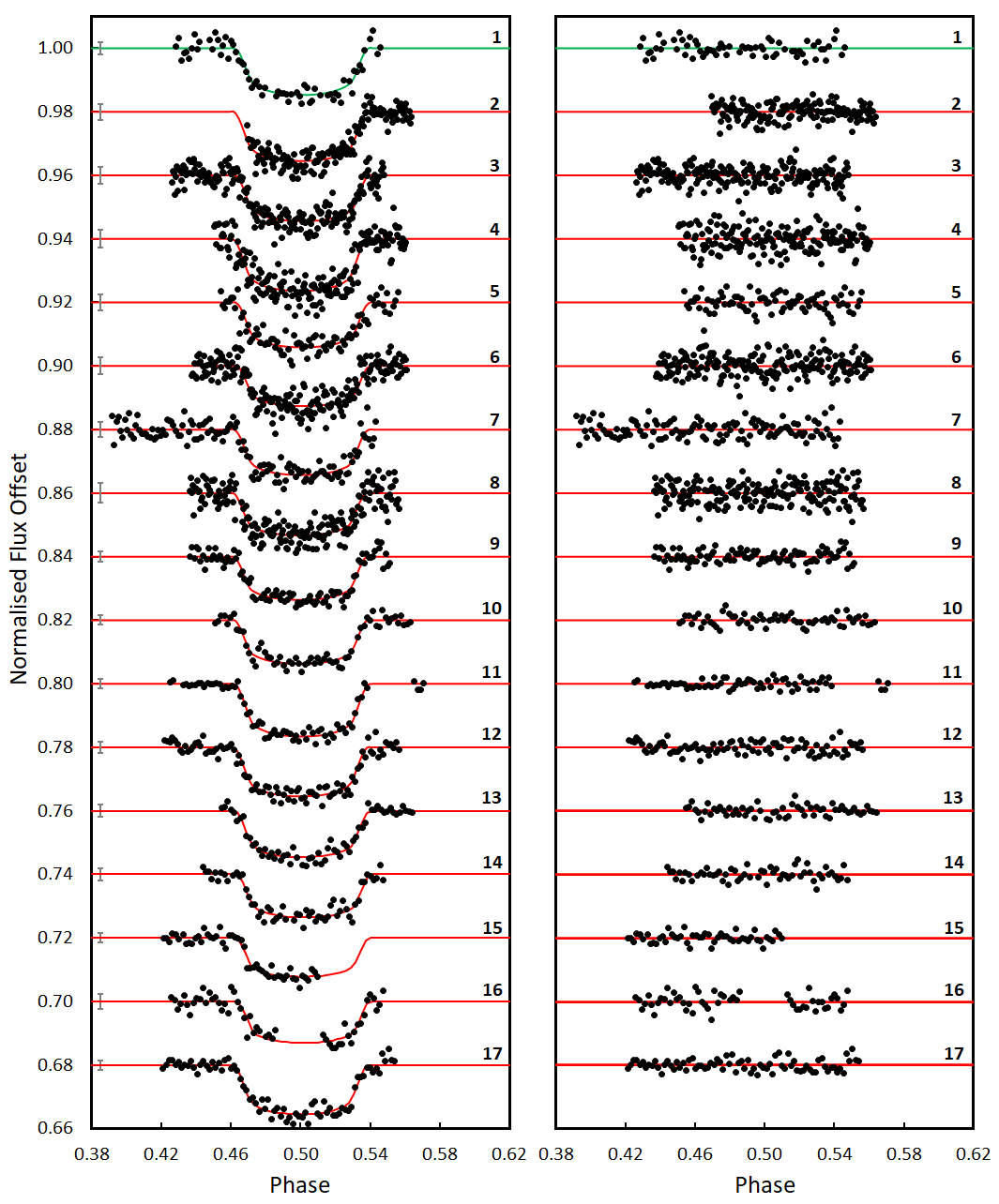}
 \caption{17 new HAT-P-23b Transit lightcurves obtained as part of this study where the transit numbering refers to the observations detailed in Table \ref{tab:Observations}.  Black dots are individual data points, the solid lines are the best fit ExofastV2 model for each transit.  Green lines represent fits to V filter observations and red lines represent fits to Rc observations.  On the left is a indication of the mean photometric uncertainty from each night's observation.}
 \label{fig:HAT-P-23b_Transits.png}
\end{figure}

\begin{figure}
 \includegraphics[width=\textwidth]{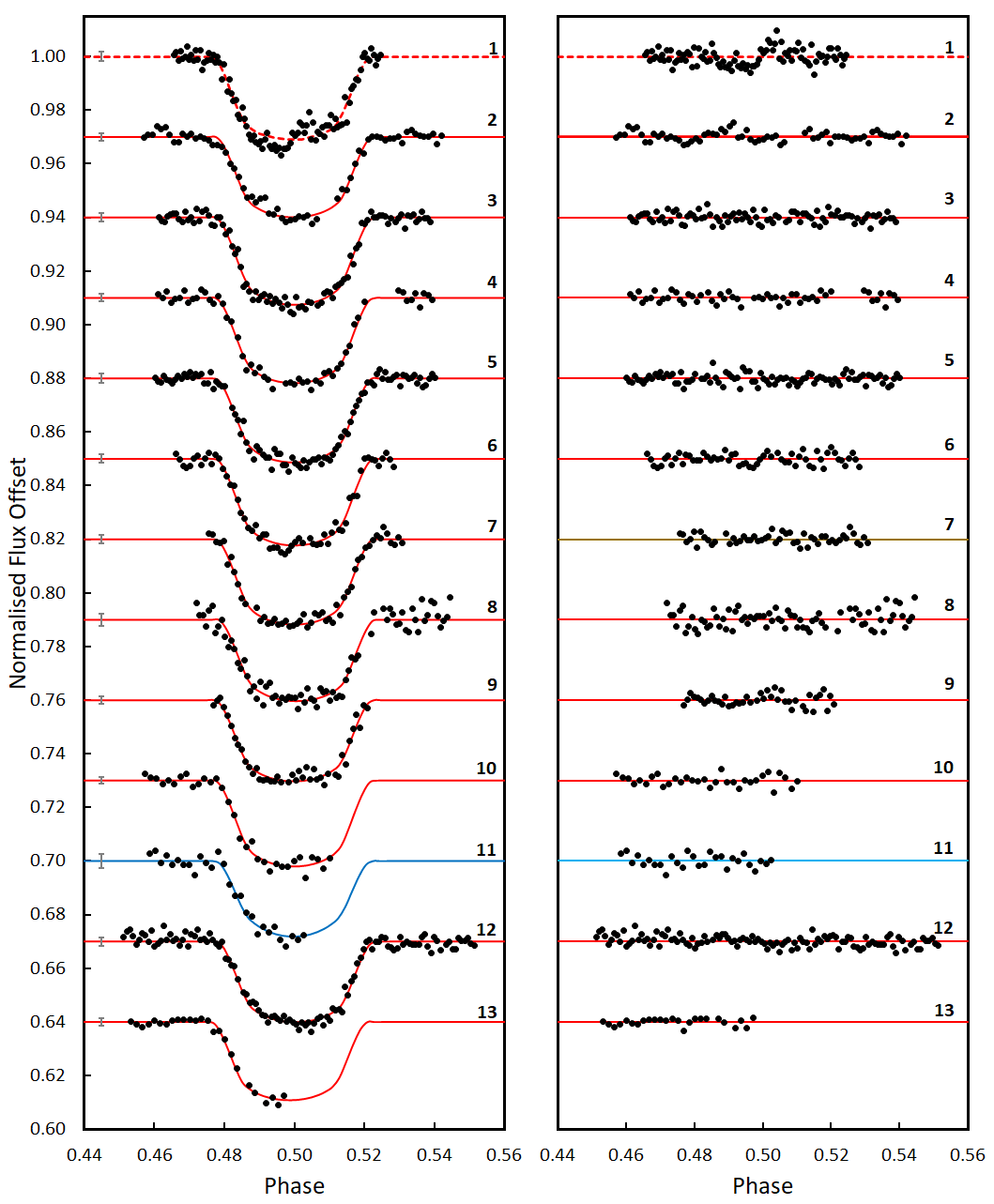}
 \caption{13 new WASP-52\,b transit lightcurves and ExofastV2 model fits obtained as part of this project.  Black dots are individual data points, the solid lines are the best fit ExofastV2 model for each transit.  Red lines represent fits to Rc observations and blue lines represent fits to B filter observations.  The numbers refer to the transits detailed in Table \ref{tab:Observations}.  On the left the mean photometric uncertainty for each observation is indicated. Only complete transits with full t1-t4 coverage were used in the updated ephemeris calculation, transit 1 (shown with a dotted model fit) was excluded from the global model fit in ExofastV2 due to the significant spot crossing event.}
 \label{fig:WASP-52b_Transits.png}
\end{figure}

To ensure that uncertainties associated with lightcurve detrending were properly accounted for in the transit fit, the airmass values and, where necessary, a pier flip time were recorded for subsequent application as part of the transit model fitting process in ExofastV2.  The lightcurve was recorded with flux normalised to unity and photometric uncertainties calculated within AIJ following Equation B1 in \cite{Collins2017}. ExofastV2 allows allows airmass and pier flips to be treated as multiplicative detrend parameters, modifying the baseline flux to achieve the best transit model fit.  In total the transit observations resulted in 1821 and 695 individual photometric measurements covering 17 and 13 transits of HAT-P-23\,b and WASP-52\,b respectively, see Figures \ref{fig:HAT-P-23b_Transits.png} and \ref{fig:WASP-52b_Transits.png}.

\subsection{Analysis}
Combined transit and radial velocity model fitting was carried out with the most recent version of ExofastV2 \citep{Eastman2019EXOFASTv2:Code} using our new transit data and previously published radial velocity (RV) measurements.  For HAT-P-23\,b 13 radial velocity measurements obtained using the HIRES instrument on the Keck telescope \citep{Bakos2011HAT-P-20b-HAT-P-23b:Planets} were used.  For WASP-52\,b 16 measurements obtained with CORALIE instrument on the ESO Euler telescope and 29 measurements obtained with the SOPHIE instrument on the 1.93m OHP telescope \citep{Hebrard2013WASP-52bPlanets} were used.  The RV data from SOPHIE was divided into separate inputs for the HE1, HE2 and HR measurements allowing different zero-points to be set for each.  The HE-RM data set was not used in our analysis.  Stellar parameters were determined through combined MESA Isochrones and StellarTracks (MIST) evolutionary track and spectral energy distribution (SED) fitting by ExofastV2 using broadband photometry obtained from published catalogues as shown in Table \ref{tab:SED photometry}, complemented with Johnson B, V and SDSS g$^{\prime}$, r$^{\prime}$ and i$^{\prime}$ photometry from release 10 of the AAVSO Photometric All-Sky Survey (APASS)\footnote{https://www.aavso.org/apass}.  Priors for the parallax were obtained from GAIA DR2 \citep{Brown2018Astrophysics2} and following the findings of \cite{Stassun2018EvidenceParallaxes}, 0.082mas was added to the reported parallax and 0.033 mas was added in quadrature to the DR2 reported uncertainty.  An upper limit for the V band extinction parameter (\textit{A}v) for each system was obtained from the Schegel dust maps\footnote{https://irsa.ipac.caltech.edu/applications/DUST/} and multiplied by 3.1 to convert to the V band extinction used in ExofastV2. This approach imposes a tight constraint on the stellar radius and the planetary parameters derived from $R_*$ \citep{Eastman2019EXOFASTv2:Code}.

All transit lightcurves, including those with partial coverage but excluding the 2013 transit of WASP-52\,b which showed a significant suspected spot crossing event, were included to produce a single global model fit in ExofastV2 for each system.  The transit mid-time for each lightcurve was allowed to vary within the global model fit and the transit mid-times were output from ExofastV2 and combined with previously published transit times to calculate a refined ephemeris for each system.  Only transits with full coverage of all transition points $\rm{t_1~to~t_4}$ were used in the ephemeris analysis.  All transit lightcurves with their best fit Exofast models can be found in Figure  \ref{fig:HAT-P-23b_Transits.png} for HAT-P-23\,b and in Figure \ref{fig:WASP-52b_Transits.png} for WASP-52\,b.

For the HAT-P-23 monitoring data a photometry template was created in AIJ incorporating stars from the UCAC4 catalogue \citep{Zacharias2013TheUCAC4} with known B, V and r$^\prime$ filter magnitudes.  Six suitable comparison stars were selected and a seventh, with B-V=1.375 similar to HAT-P-23 (B-V=1.301), was used as a check star.  AIJ does not take into account catalogue magnitude uncertainties so we report the uncertainties for the calculated magnitudes based on the standard error in the measurements of the check star.    

\section{Results}\label{Results}

\subsection{HAT-P-23b}
\begin{figure*}
 \includegraphics[width=\textwidth]{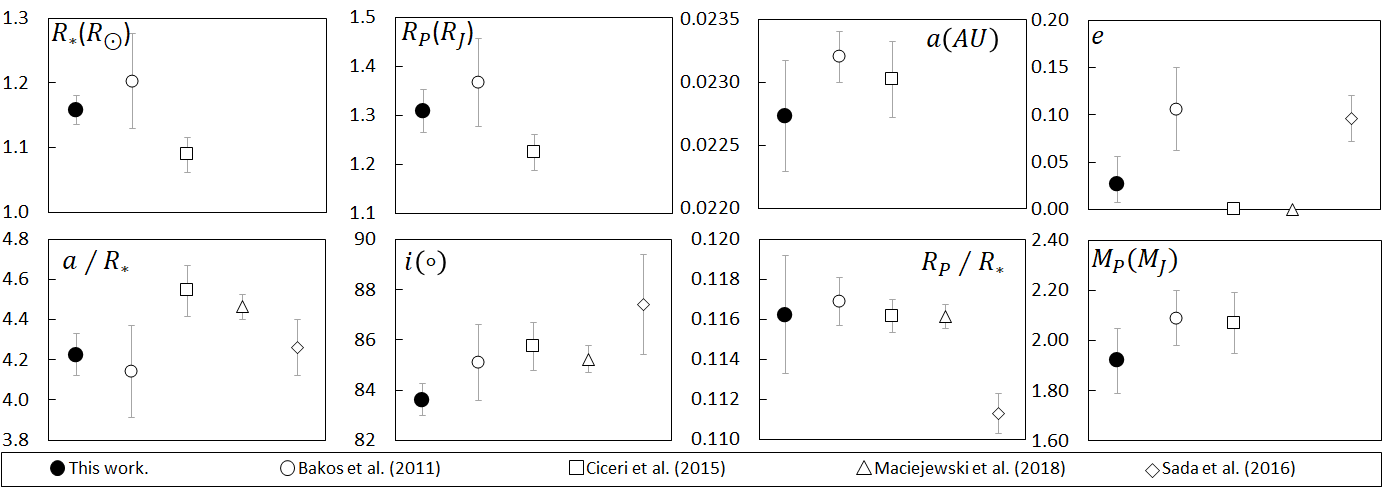}
 \caption{HAT-P-23\,b results from ExofastV2 model analysis for our new transits (filled circles) compared with results from other published works where values are available (open symbols).  Open circles are from \protect\cite{Bakos2011HAT-P-20b-HAT-P-23b:Planets},  squares are from \protect\cite{Ciceri2015}, triangles are \protect\cite{Maciejewski2018Planet-starB} and diamonds are the values from \protect\cite{Sada2012}. }
 \label{fig:HAT-P-23b_Exofast_Outputsv2.png}
\end{figure*}

At discovery \cite{Bakos2011HAT-P-20b-HAT-P-23b:Planets} found a marginally significant orbital eccentricity of $0.106\pm\,0.044$.  The subsequent study by \cite{Sada2016} also found an orbital eccentricity of $0.096\pm\,0.024$, however using secondary occulatation measurements \cite{ORourke2014WarmHat-p-23b} determined that the orbit was probably circular, a result supported by subsequent RV studies \citep{Bonomo2017ThePlanets}.  Subsequent studies of HAT-P-23\,b made by \cite{Ciceri2015} and \cite{Maciejewski2018Planet-starB} assumed a circular orbit.  \cite{Maciejewski2018Planet-starB} reported that the increased value they found for the scaled system parameter $a/R_*$, over that seen by \cite{Bakos2011HAT-P-20b-HAT-P-23b:Planets} and \cite{Sada2016} is due to the adoption of a non zero eccentricity in those studies.  To ensure we did not bias our results we left the orbital eccentricity as a free parameter in our ExofastV2 fits.  This resulted in $e = 0.027^{+0.029}_{-0.019}$, consistent with a circular orbit within $1.4\sigma$.  We also calculated the same model with a eccentricity forced to zero finding exactly the same results for the scaled $a/R_*$ parameter as we found with the eccentricity left as a free parameter indicating this unlikely the cause of the difference seen by \cite{Maciejewski2018Planet-starB}.  Our result for $a/R_*$ is consistent with that determined by \cite{Bakos2011HAT-P-20b-HAT-P-23b:Planets} and \cite{Sada2016}, Figure \ref{fig:HAT-P-23b_Exofast_Outputsv2.png}.

From our global model we find the stellar radius and mass of HAT-P-23 to $1.157^{+0.023}_{-0.022}\rm{R}_{\odot}$ and $1.063^{+0.063}_{-0.060}\rm{M}_{\odot}$ which is 3.8\% and 5.9\% smaller respectively than found by \cite{Bakos2011HAT-P-20b-HAT-P-23b:Planets}.  Our results are also 6.2\% larger and  3.7\% smaller respectively than the only other published stellar mass and radius values  by \cite{Ciceri2015}.  As a result of our new stellar radius measurement we find the radius of HAT-P-23\,b to be $1.308^{+0.44}_{-0.43}\rm{R}_{\rm{J}}$.  We find $\rm{a/R_*} = 4.22^{+0.11}_{-0.10}$, consistent with the previous studies which assumed a non zero eccentricity \citep{Bakos2011HAT-P-20b-HAT-P-23b:Planets,Sada2016}  and approximately 6\% smaller than that found by studies which fixed orbital eccentricity to zero \citep{Ciceri2015,Maciejewski2018Planet-starB}. Stellar and planetary parameters are shown in Figure \ref{fig:HAT-P-23b_Exofast_Outputsv2.png}  and Table \ref{tab:Exofast outputs}.

\begin{figure*}
 \includegraphics[width=1.0\textwidth]{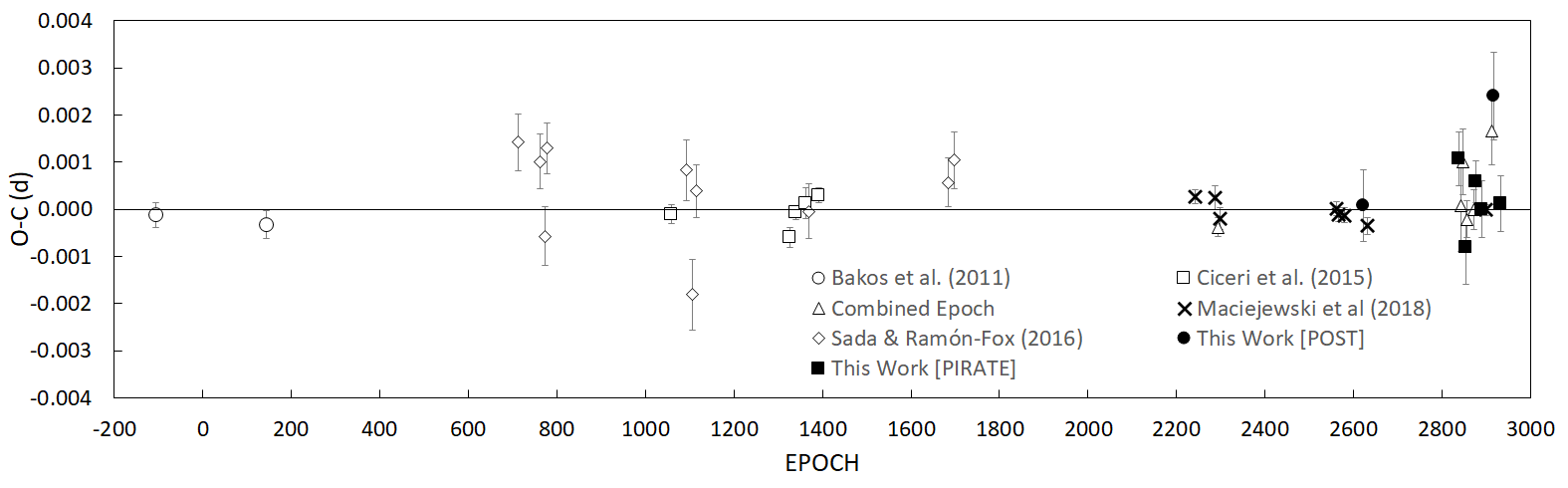}
 \caption{Transit observed - calculated (O-C) times for HAT-P-23\,b from the literature and this work calculated using our new ephemeris. Transit times for complete transits where lightcurve data was available were recalulated \citep{Bakos2011HAT-P-20b-HAT-P-23b:Planets,Ciceri2015,Maciejewski2018Planet-starB}.  Other transit times were taken from the literature and converted to BJD$_{TDB}$ as required. The mean value derived for simultaneous observations of the same transit are shown as open triangles.}
 \label{fig:HAT-P-23b_Ephemeris.png}
\end{figure*}

\begin{figure*}
 \includegraphics[width=1.0\textwidth]{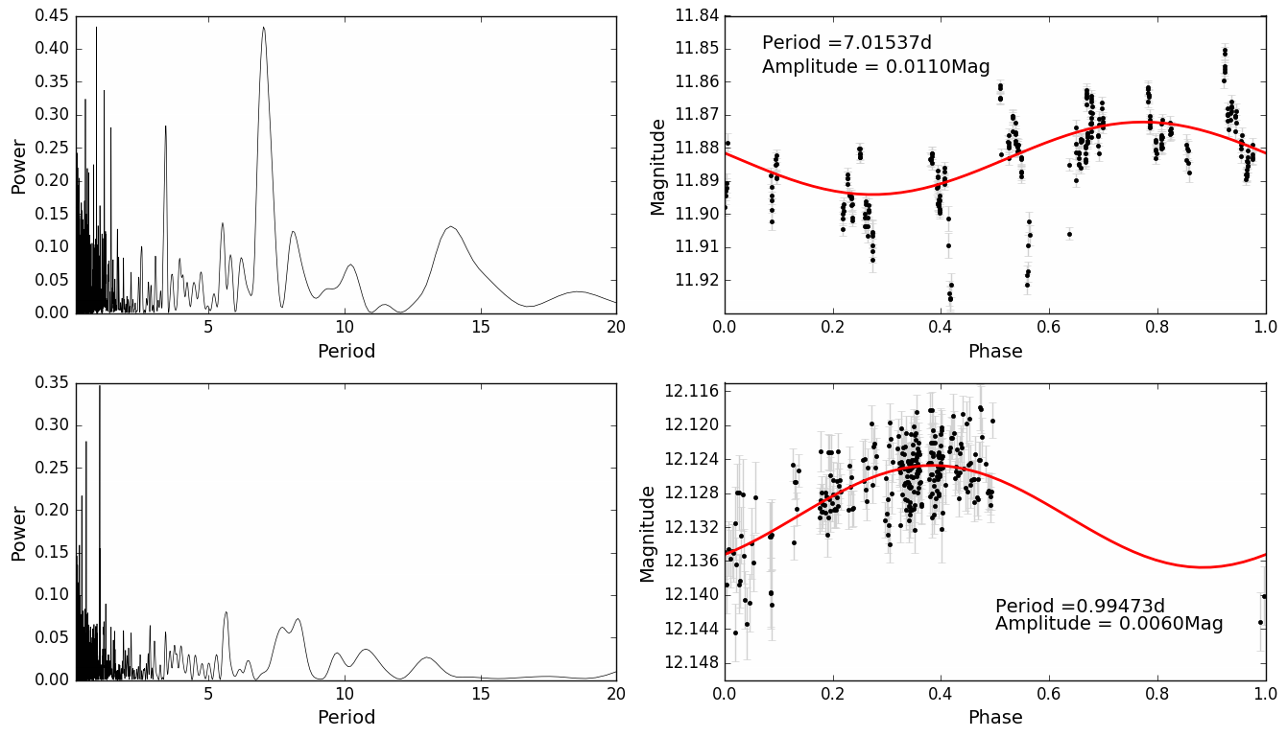}
 \caption{Lomb Scargle periodograms and phase folded results showing the detected rotation period for HAT-P-23 (top) and for the check star (bottom).  The period region shown is between 0.1d and 20d, no periods were detected outside of this range.  The period detected in the check star is the one day observing cadence (and it's aliases).}
 \label{fig:HAT-P-23b_Monitoring.png}
\end{figure*}

To calculate an updated ephemeris for HAT-P-23\,b the transit mid times for all complete transits were combined with those from the literature.  Full transit lightcurves published by \cite{Ciceri2015} and \cite{Maciejewski2018Planet-starB} which were available on-line were reanalysed in ExofastV2 in exactly the same way as for our new transit data.  We generally reproduced the published mid times within the measurement uncertainties.  We included only the R filter data from the simultaneous transit measurements made by \cite{Ciceri2015} using the BUSCA instrument.  Where the lightcurve data was not available, transit mid times from other publications were taken from the literature corrected to $\textrm{BJD}_{\rm{TDB}}$ as required.  The resulting transit mid-time data set contained a number of simultaneous observations of the same transit either by the same or by different authors.  In these cases a simple mean value for the transit mid-time was used.  The transit mid-times from our newly obtained lightcurves can be found in Table \ref{tab:HAT-P-23b transit times}, and a full list of the transit times used in the ephemeris calculation is available in the supplementary material.  The new linear ephemeris found using this data set is

\begin{equation}\label{eq:1}
\begin{aligned}
T_{c} (BJD_{TDB}) = 2454852.265165(120) + 1.212886457(54)\rm{d} \times E 
\end{aligned}
\end{equation}
where E is the epoch of observation and values in brackets are the uncertainties relative to the last digit.  The $\chi_{\nu}^2$ of the linear fit is 2.08 and the Bayesian Information Criterion (BIC) is 33.01.We also calculated a quadratic ephemeris of the form;  
  
\begin{equation}\label{eq:2}
T_{c} (BJD_{TDB}) = T_0  + P_{\textrm{orb}} \times E + 0.5(\delta P_{\textrm{orb}}/\delta E) \times E^2  
\end{equation}
resulting in $\chi_{\nu}^2 = 2.20$ and BIC = 37.75.  Therefore we find no departure from a linear ephemeris for HAT-P-23b, in agreement with the results from \cite{Maciejewski2018Planet-starB}.  The timing residuals to our linear ephemeris are shown in Figure  \ref{fig:HAT-P-23b_Ephemeris.png}
\paragraph*{}
The monitoring data obtained for HAT-P-23b was analysed to look for periodic variations due to star spots on the rotating stellar surface.  A Lomb-Scargle periodogram analysis over the period range of 0.1d to 100d reveals a 7.015d period with an amplitude of 1.1\%, substantially less than the 3\% seen by \cite{Sada2016} (\ref{fig:HAT-P-23b_Monitoring.png}). \cite{Moutou2011Spin-orbitHAT-P-23b} measured HAT-P-23b’s rotational velocity as $\rm{7.8\pm1.6km s^{-1}}$ which, given our newly determined stellar radius, results in an equatorial rotation period of $\rm{7.50\pm1.55}$ days, consistent with our measured period of variation.  We therefore attribute the variability seen to the rotation period of HAT-P-23 at the (unknown) latitude of the surface spots.

\subsection{WASP-52b}
As with HAT-P-23\,b a joint transit and RV fit was undertaken using all new full and partial transit data sets, except transit number 1 which exhibits a large suspected spot crossing event.  Again to avoid bias through fixing model parameters we left orbital eccentricity as a free parameter in the ExofastV2 fits, resulting in $e = 0.05^{+0.033}_{-0.03}$, within $1.6\sigma$ of a circular orbit, thus the eccentricity found is consistent with zero.

\begin{figure*}
 \includegraphics[width=1.0\textwidth]{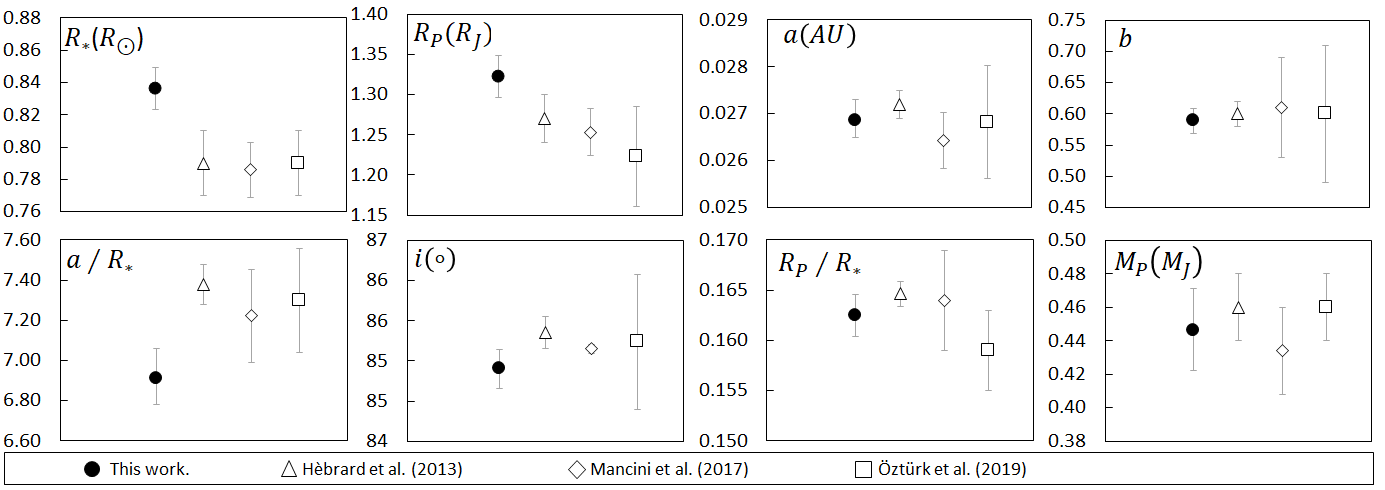}
 \caption{WASP-52\,b key system parameters from both this work (filled circles) compared with those from \protect\cite{Hebrard2013WASP-52bPlanets} (open triangles), \protect\cite{Mancini2017OrbitalSystem} (open diamonds) and \protect\cite{Ozturk2019NewWASP-52b} (open squares).  Semi-major axis and impact parameter shown for the \protect\cite{Ozturk2019NewWASP-52b} results were calculated from published parameters as was the scaled $R_P/R_*$ and impact parameter for \protect\cite{Mancini2017OrbitalSystem}}
 \label{fig:WASP-52b_Results_Comparison.png}
\end{figure*}

We find a stellar radius of $0.834\pm\,0.013\rm{R}\odot$ which is 5.6\% larger than that found by \cite{Hebrard2013WASP-52bPlanets} and 6.1\% larger than found by \cite{Mancini2017OrbitalSystem}.  This larger stellar radius value results in a planetary radius of $1.319^{+0.026}_{-0.027}\rm{R}_{\rm{J}}$, 3.9\% and 5.3\% larger respectively than found in those studies.  The larger value we find for the stellar radius results from the use of an extinction corrected SED model with Gaia parallax to constrain the system distance.  We find the distance to WASP-52\,b is $173\pm\,1.6\,\rm{pc}$, 24\% greater than the $140\pm\,20\,\rm{pc}$ found by \cite{Hebrard2013WASP-52bPlanets}.  Our result for the semi-major axis remained within 2\% of the results from previous studies so our derived scaled system parameter $\rm{a/R}_*$ is smaller than previous results by 5.8\% and 3.8\% respectively.  Other parameters determined are consistent with previously published results.  Our results for key parameters are presented in Figure \ref{fig:WASP-52b_Results_Comparison.png} along with previously published values.  Our derived effective temperature of $5039^{+40}_{-39}$K is consistent with \textrm{$T_{\rm eff}$}$ = 5000\pm\,100$K from \cite{Hebrard2013WASP-52bPlanets}, though with improved uncertainties.  The age reported for WASP-52 varies widely in the literature from $0.4^{+0.3}_{-0.2}$Gyr \citep{Hebrard2013WASP-52bPlanets} to $9.4^{+4.7}_{-4.3}$Gyr in \cite{Mancini2017OrbitalSystem}.  We find an age of $8.5^{+3.7}_{-4.6}$Gyr agreeing within 1$\sigma$ with \cite{Mancini2017OrbitalSystem}.

The transit mid times for all 9 new transits with complete coverage (excluding archive transit number 1) were combined with those from the literature from \cite{Hebrard2013WASP-52bPlanets}, \cite{Swift2015MiniatureResults}, \cite{Bruno2018StarspotWASP-52b}, \cite{Mancini2017OrbitalSystem}, \cite{Ozturk2019NewWASP-52b} and \cite{Baluev2019HomogeneouslyWASP-4}.  Table \ref{tab:WASP-52b transit times} lists the new transit times determined from our observations, the full table of transit times used in the ephemeris calculations can be found in the supplementary material to this paper.  The transit times calculated by \cite{Baluev2019HomogeneouslyWASP-4} make extensive use of observations recorded on the Exoplanet Transit Database; only those results classified as being from "high quality" lightcurves by those authors were included in our analysis.  The transit times published by \cite{Mancini2017OrbitalSystem} have extremely small uncertainties and including these times in our fit was found to increase the $\chi_{\nu}^2$  of the ephemeris.  The authors noted that their linear ephemeris had a $\chi_{\nu}^2$ of 8.98 and they subsequently multiplied the uncertainties on their linear ephemeris by this value.  We applied this same multiplier to their individual transit mid-time uncertainties prior to calculating our new ephemerides.  We first calculated a linear ephemeris resulting in a fit with  $\chi_{\nu}^2 = 1.58$ and BIC = 59.41.

\begin{equation}\label{eq:3}
\begin{aligned}
T_{c}(BJD_{TDB}) = 2455793.681914(141) + 1.749781099(126)\rm{d} \times E 
\end{aligned}
\end{equation}
A quadratic ephemeris of the form from Equation \ref{eq:2} was also modelled resulting in a slightly lower $\chi_{\nu}^2 = 1.51$ and BIC = 57.88.  

\begin{equation}\label{eq:4}
\begin{aligned}
T_{c}(BJD_{TDB}) = 2455793.680977(141) + 1.749783241(177)\rm{d} \times E\\
            +~(-2.14\pm0.24) \times 10^{-9}\rm{d} \times E^2
\end{aligned}
\end{equation}

\begin{figure*}
 \includegraphics[width=\textwidth]{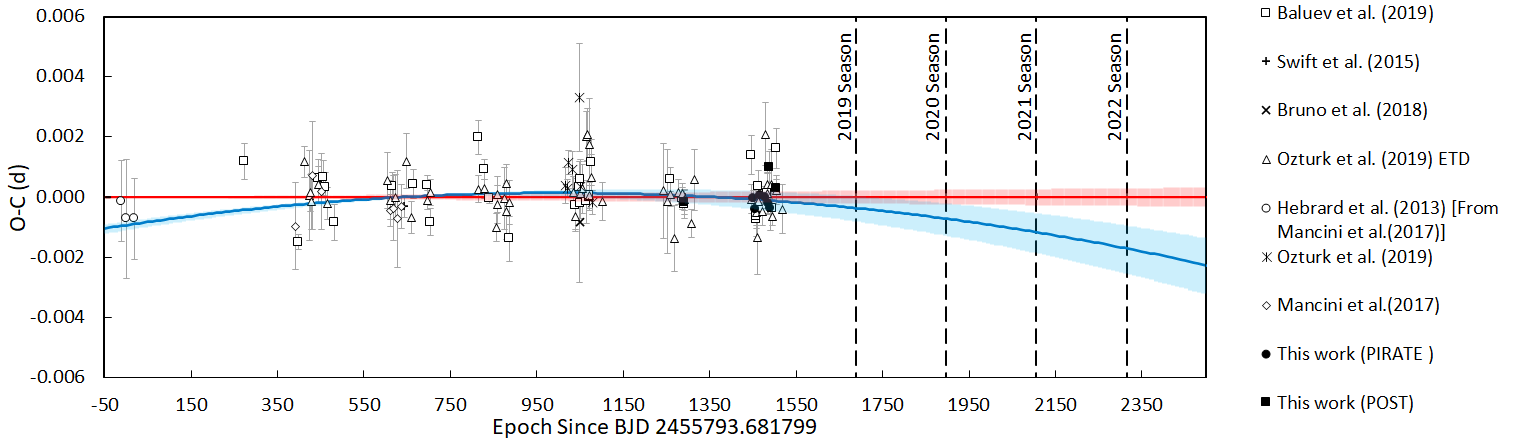}
 \caption{Predicted transit times for WASP-52\,b based on both the linear and quadratic ephemerides shows that it should be possible to clearly differentiate between the two models with data from the 2021 observing season. }
 \label{fig:WASP-52b_Ephemeris_Prediction.png}
\end{figure*}

 Using \cite{Mancini2017OrbitalSystem}'s originally published uncertainties results in the linear model being preferred over the quadratic result while increasing the $\chi_{\nu}^2$ values to 2.3 for the linear model and 2.8 for the quadratic model.  Thus the small published uncertainties on these results appear to have masked previous detection of a non-linear ephemeris for WASP-52b.  Excluding the transit times reported by \textbf{}\cite{Mancini2017OrbitalSystem}, including their recalculation of the three transit times from \cite{Hebrard2013WASP-52bPlanets}, produces increased $\chi_{\nu}^2$ for both ephmerides but retains a preference for the quadratic model. Finally we removed our new transit times and find the quadratic model is still preferred, though again with an increased $\chi_{\nu}^2$.  In our quadratic ephemeris we found that the values for the quadratic term between 1.90 and 2.38 resulted in the same $\chi_{\nu}^2$ values (measured to 2 decimal places), therefore we increased our formal uncertainty on the period derivative to reflect this range of possible values.  The quadratic term $\frac{\delta P}{\delta E} = (-2.14\pm\,0.24)\times10^{-9}$ can be used to calculate the period derivative of

\begin{equation}\label{eq:5}
\begin{aligned}
\frac{\delta P}{\delta t} = \frac{1}{P}\frac{\delta P}{\delta E} = (-1.22\pm0.14)\times10^{-9} = -38.6\pm4 {\rm ms~yr^{-1}}
\end{aligned}
\end{equation}
This is 34\% larger than seen for the largest secure period change detected for WASP-12\,b \citep{Yee2019TheDecaying}.  We discuss the quadratic ephemeris in Section \ref{Wasp-52b_Discussion}.  Figure \ref{fig:WASP-52b_Ephemeris_Prediction.png} shows the O-C values.
\paragraph*{}
WASP-52 is an active star with many spot crossing events previously recorded, e.g. \cite{Mancini2017OrbitalSystem, Kirk2016TransmissionSurface}.  No suspected spot crossing events are visible in our transits observed during the 2018 season however transit number 1 obtained in 2013 shows a large brightening during the transit.  This was the same season as the observations made by \cite{Mancini2017OrbitalSystem} who detected spot crossing events at 5 of their 8 observation epochs, thus we interpret this event as a significant spot crossing.  Figure \ref{fig:WASP-52b_Transits.png} shows how ExofastV2 can significantly underestimate the transit depth, and potentially transit time, where a large spot crossing occurs therefore this transit was excluded from the ExofastV2 model fit and ephemeris calculation for WASP-52\,b.  The final derived system parameters and uncertainties are detailed in Table \ref{tab:Exofast outputs}.

\section{Discussion}\label{Discussion}

\subsection{HAT-P-23b}
Previous results for the radius of HAT-P-23b have varied from an inflated $1.368\pm\,0.09 \rm{R_J}$ \cite{Bakos2011HAT-P-20b-HAT-P-23b:Planets} to the $1.6\sigma$ smaller radius of $1.224^{+0.036}_{0.007} \rm{R_J}$ \citep{Ciceri2015}.  Our derived radius of $1.308^{+0.044}_{-0.043}\rm{R_J}$ is between these two values with a slightly smaller mass.  This places HAT-P-23\,b very close to two other inflated hot Jupiters, TrEs-3\,b \citep{ODonovan2007} and WASP-135\,b \citep{Spake2016WASP-135b:Star} in terms of of planetary radius,  mass and semi major axis.  It is notable though that HAT-P-23\,b receives almost twice the incident flux resulting in an equilibrium temperature that is 300-400K greater than for the other two planets, Table \ref{tab:H23_T3_W135_Comp}. A strong dependency on radius with incident flux was found by \cite{Weiss2013TheFlux} who derived an empirical relationship for planetary radius dependent on mass and incident flux noting that for planets with $\rm{M_P > 150M_{\oplus}}$ the radius depends only very weakly on mass, thus the incident flux is the most important factor affecting planetary radius in this mass range.  Using this empirical relation and our new planetary mass and incident flux we calculate a theoretical radius for HAT-P-23\,b of $1.355\pm\,0.008 M_{\rm{J}}$ which is larger than the radius we measure, closer to that determined by \cite{Bakos2011HAT-P-20b-HAT-P-23b:Planets}, though our measured and the theoretical radii are close to agreement at $1\sigma$.  It is likely that factors other than just incident flux such as core mass fraction, atmospheric composition or age play an important role in determining the radius of HAT-P-23\,b.  The determination of precise planetary and host star parameters for known hot Jupiters along with the discovery of new examples will allow comparative planetology to tease out the relative importance of the factors affecting the radii of hot Jupiters.

The value we find for the eccentricity of $e = 0.027^{+0.029}_{-0.019}$ is consistent with zero and thus in agreement with the findings from secondary eclipse measurements and revised RV analysis \citep{ORourke2014WarmHat-p-23b,Moutou2011Spin-orbitHAT-P-23b}.  Figure \ref{fig:HAT-P-23b_Exofast_Outputsv2.png} shows previous studies treating eccentricity as a free parameter yield a smaller scaled system parameter $\rm{a/R_*}$ than studies where the eccentricity is fixed to zero.  This biases results towards larger values of $\rm{a/R_*}$ where a circular orbit assumption is made, however we see no difference in $\rm{a/R_*}$ between our free eccentricity and circular models. 
\paragraph*{}
We have presented an updated ephemeris for HAT-P-23\,b based on our new observations and remodelling of those published observations where the lightcurve data is available.  We have included only transits with full coverage of both the ingress and egress phases in our ephemeris calculation and have mean combined simultaneous transit times for six epochs where the same transit was observed by multiple observers.  We find a linear ephemeris is the best fit to the available data with a $\chi_{\nu}^2 = 2.02$.  Such $\chi_{\nu}^2$ values are not uncommon in exoplanet ephemeris results that combine data sets from multiple observers, for example \cite{Southworth2019TransitSystem}.
\paragraph*{}
Motivated by the detection of a 3\% brightness variation by \cite{Sada2016} during their observing campaign between 2012 and 2014, we undertook photometric monitoring of HAT-P-23 over a 78-day period in 2018. We detected a periodic variation of 7.015 days (\ref{fig:HAT-P-23b_Monitoring.png}) which we attribute to the rotation of spots on the surface of HAT-P-23b. The amplitude of the variation seen was 1.1\%, significantly less than that seen by \cite{Sada2016}, suggesting that the activity cycle of HAT-P-23 has declined since their observations were made.

\subsection{WASP-52b}\label{Wasp-52b_Discussion}
Long term monitoring of known transiting hot Jupiters has revealed several systems suspected of exhibiting a non-linear ephemeris such as WASP-12\,b \citep{Maciejewski2016}, WASP-4b \citep{Bouma2019WASP-4bMission} and Kepler-1658 \citep{Chontos2019TheExoplanet}.  Orbital decay is postulated as the cause of the declining orbits while apsidal precession is an alternative explanation and it is difficult to distinguish between these explanations with the available timeline of observations.  This is the motivation for extending the timeline of precise transit mid-time observations for short period hot Jupiter systems using readily available small aperture telescopes. 

We have presented an updated ephemeris for WASP-52\,b using our new observations obtained with multiple small aperture telescopes combined with previously published results and those available from the Exoplanet Transit Database (ETD) and find a quadratic ephemeris provides the best fit to the available data.  The $\Delta\chi_{\nu}^2 = 0.07$ between the linear and quadratic ephemerides is small and has not been reported in previous analysis of transit times for WASP-52\,b, \citep{Mancini2017OrbitalSystem, Ozturk2019NewWASP-52b, Baluev2019HomogeneouslyWASP-4}.  Extrapolating both ephemerides shows that with timing measurements from the 2021 observing season it should be possible to determine if the linear or quadratic model is correct, (Figure \ref{fig:WASP-52b_Ephemeris_Prediction.png}).  Below we consider some possible causes for the period change implied by the quadratic ephemeris model result.

\subsubsection{Orbital Decay}
Our result for the period change implies a decay time (based on linear decay to $a = 0$) of ~4.1Myr, approximately 0.05\% of the derived age for WASP-52.  Following equation 14 from \cite{Patra2017} and using $\delta P / \delta t = -1.22\pm\,0.14\times10^{-9}$ we calculate the modified tidal quality factor to be $Q^{'}_{*} = 1.1\pm0.2\,\times10^{3}$.  This is at least 2-3 orders of magnitude smaller than typically expected \cite[][and references therein]{Patra2017}, ruling out tidal decay as a potential cause of the observed period change.  

\subsubsection{Apsidal Precession}
Apsidal precession requires the planet has an eccentric orbit and with a short 1.75d period, a method of maintaining that eccentricity is not clear.  Assuming a more likely maximum value for $Q^{'}_{*}$ of $10^5$ the tidal circularisation timescale, $\tau_{\rm circ}$, is just ~5Myr, significantly less than the $8.5^{+3.7}_{-4.6}$Gyr age of the system.  Our analysis finds no significant evidence for a departure from a circular orbit, however we note a small eccentricity is possible and additional radial velocity monitoring or secondary occultation observations would be required to confirm the existence, or otherwise, of a small orbital eccentricity.

\subsubsection{Applegate Effect}
In the Applegate effect \citep{Applegate1992}, changes in the stellar quadrupole moment drive angular momentum transfer within the star over the magnetic cycle period of the star.  If angular momentum is transferred from the core to the stellar envelope the star will become more oblate.  An orbiting planet period change due to the Applegate effect has been considered by \cite{Watson2010}.  To maintain a constant angular momentum in the changing gravitation potential as the star becomes more oblate the orbital speed must increase and therefore the period decrease. The opposite is true if the oblateness of the star decreases, the period will increase, giving rise to a sinusoidal change in the orbiting planet's transit times.  Following Equation 13 from \cite{Watson2010}, setting the time-scale over which the quadrupole changes occur to either the approximate 7.5 years duration of WASP-52b observations, or twice this value, the amplitude of period change is 0.244 or 0.59 seconds respectively.  This equates to a period change of $37\pm\,10{\rm ms~yr^{-1}}$ for the 7.5 year quadrupole change or $53\pm\,14{\rm ms~yr^{-1}}$ if the cycle is 15 years, consistent with the detected period change value of $-38.6\pm\,4{\rm ms~yr^{-1}}$ so orbital period changes due to the Applegate effects cannot be ruled out. For a given set of stellar parameters the magnitude of the O-C value arising from the Applegate effect is sensitive to both the stellar rotation period and the assumed modulation period, with longer a rotation period reducing the O-C value and longer modulation periods increasing it.  The stellar rotation period from \cite{Hebrard2013WASP-52bPlanets} is dependent on the poorly constrained inclination of the system. Based on spot crossing events \cite{Mancini2017OrbitalSystem} derived a rotation period of $\rm{15.3\pm1.96d}$ which yields period change values of $28\pm\,5{\rm ms~yr^{-1}}$ for an assumed 7.5 year modulation period. The modulation period is therefore main unknown parameter and will require further observations to determine.

\subsubsection{Distant Perturber}
\cite{Hebrard2013WASP-52bPlanets} identified a possible long term RV trend of $+40\rm{ms^{-1}}$ over a period of 15 months which they concluded could be due to a distant companion or a systematic effect resulting from combining multiple data sets.  A companion in the range of $10-15\rm{M_J}$ out to 6AU was ruled out by a lack of transit timing variation signal \citep{Swift2015MiniatureResults}.  In their  re-analysis of publicly available data from the ETD \cite{Baluev2019HomogeneouslyWASP-4} also find no periodic variation attributable to TTVs.  A Lomb-Scargle period analysis was undertaken of the combined data set with our new observations finding no evidence of periodicity resulting from TTV's. The acceleration away from Earth of $+40{\rm ms^{-1}}$ over 15 months would produce a radial velocity acceleration $\dot{\nu}_r \approx 0.09{\rm ms^{-1}d^{-1}}$ resulting in a positive period derivative of approximately $\dot{P} \approx \dot{\nu}_rP/c \approx 5\times10^{-10}$, of opposite sign and an order of magnitude smaller than seen.  The measured $\dot{P}$ of $-39{\rm ms~yr^{-1}}$ would require a $\dot{\nu}_r \approx 0.2{\rm ms^{-1}d^{-1}}$, which would be readily detectable in the available radial velocity measurements.

\subsubsection{Star Spots}
The impact of star spots on transit time measurements has been considered by previous authors \citep[E.g.][]{Oshagh2013, Ioannidis2016}.  \cite{Barros_2013_WASP-10b} identified spot transits as the cause of TTVs in WASP-10\,b.  \cite{Ioannidis2016} showed that timing from higher precision lightcurves is more affected by stellar spots, and they  define the transit SNR (TSNR) as the  transit depth divided by the out of transit light curve noise level.  They demonstrated that for $\rm{TSNR} \lesssim 15$ systematic timing errors introduced by spot crossing events are not distinguishable from the random noise.  They also showed that the magnitude of the effect of spots on timing measurements is determined by the spot longitude on the stellar surface, peaking at $\lambda = \pm70$\textdegree, reducing the closer a spot is to the stellar limb (as the size of the projected spot is reduced) and becoming zero at the centre of the stellar disk.  Therefore a spot crossing will induce a positive time variation where it occurs before the transit mid-time and a negative variation where it occurs after transit mid-time, leading to a transit time variation effect than can mimic that caused by a companion planet.  For a well aligned and prograde planetary orbit such as that for WASP-52\,b \citep{Hebrard2013WASP-52bPlanets, Mancini2017OrbitalSystem} this TTV mimicking effect would result in detectable variations, or at least an increase in the spread of O-C measurements if the sampling is poor, rather than the declining period we see in the ephemeris of WASP-52\,b.  

The approximately 2.7\% deep transits of WASP-52b means that an out of transit noise level of 0.0018 or less is required to achieve $TSNR>15$, a level of photometric precision marginally achieved in only a small number of our new WASP-52\,b transits used in the ephemeris calculation.  Therefore we would not expect our transit mid-time measurements to be systematically affected by spot crossing events.  While no spot crossing events were positively identified in the new photometry obtained in the 2017-18 season the 2013 archive observation from the UK shows a clear brightening during transit.  This observation was obtained during the same period as those by \cite{Mancini2017OrbitalSystem} who positively detected spot crossing events at 5 of their 7 epochs with full transit coverage.  The absence of spot crossing events in our 2017-18 transits could indicate that the photometric precision achieved is insufficient to resolve the spot crossings or that since the 2012-2014 seasons WASP-52's stellar cycle has either quieted or the spot latitude has migrated away from the transit chord.  If the spots have ceased or migrated to a different latitude from the transit chord this would be apparent in photometric monitoring of WASP-52.  Stellar spots occurring at a latitude different to the transit chord would be visible as stellar variability and potentially detectable as a correlation between stellar flux and transit depth \citep[E.g.][]{Kirk2016TransmissionSurface}.  If WASP-52 is in a reduced activity phase of its stellar cycle we would not expect to see either effect.  

Unfortunately concurrent photometric monitoring of WASP-52 over the 2018 season is not available; instead the stellar magnitude was measured using out of transit data from four transits observed using PIRATE with sufficient unbroken data available on both sides of the transit (transits 3 to 6 in Table \ref{tab:Observations}).  As for HAT-P-23 a photometry template was generated using comparison stars from the UCAC4 catalogue.  Eight stars were used in the range r$^{\prime} = 13.00-14.90$ mag.  WASP-52 itself has no B, V or r$^{\prime}$ photometry in the UCAC4 catalogue so $B-V = 0.9$ was taken from \cite{Hebrard2013WASP-52bPlanets} and a check star of similar $B-V = 0.86$ was selected from UCAC4.  Although this data set contained data from only 4 observations a variation of $\Delta Rc = 0.03$ mag, similar to the WASP-52b transit depth, was seen.  

The lack of transit spot crossing events in the 2017-18 season along with the brightness variation we see in WASP-52 indicate WASP-52 is still active but that the activity latitude has migrated away from the transit chord.    Observations in future seasons, ideally with a B filter to increase spot contrast,  would be required to follow the stellar activity cycle and determine its periodicity.  This result shows how observations obtained using small aperture telescopes, such as those used in this study, can provide monitoring of transiting systems to identify activity cycles and help advise the planning of transmission spectroscopy observations which are complicated by activity on the host star.

\section{Observatory Performance and Capability}\label{Observatory}
We now consider the performance and capabilities of the small small aperture telescopes used in this study to undertake long duration transit follow up campaigns.  In observations obtained with PIRATE before 17th July 2018 and for observations made with POST the individual exposure start times were recorded in the FITS headers using the PC clock controlled using the separate application Dimension4 \footnote{(http://www.thinkman.com/dimension4/)} to continually update the control PC clocks from an Internet time server.  During an upgrade carried out 17th - 20th July 2018 GPS time controllers were installed at both PIRATE and COAST observatories with camera shutter timing measurement to record the exposure start and end times to sub-millisecond accuracy.  Shortly before this upgrade the imaging process for the observations in this study was changed from 2x2 on-chip binning  to 1x1 un-binned imaging.  The observations of HAT-P-23b either side of this period allow a comparison of the timing precision achievable before and after installation of the GPS timing control.  The O-C results for HAT-P-23b show a three fold reduction in the measurement scatter after the upgrade date with the RMS scatter reducing from 93.41s to 33.48 seconds.  At the same time the mean individual transit mid-time uncertainties were reduced by 17\% from 62.6s to 51.8 seconds, likely due to the higher precision photometry produced from the change to 1x1 binning.  Over the same period the transit O-C RMS scatter determined from POST observations remain unchanged at 113.7 seconds with an individual transit mid-time uncertainty mean of 70.3s.  The observations of WASP-52b obtained with PIRATE all occurred after the GPS installation and were all made unbinned, resulting in an O-C RMS scatter of 16.4s and a mean transit mid-time uncertainty of 25.9s (excluding the transit of 10th October 2018 without out of transit coverage).
 
It is not possible to separate the timing improvement due to the GPS timing implementation from the change to 1x1 binned observations though it is expected both will have contributed to the reduced spread in O-C measurements seen.  New planned observations of transit targets with both 1x1 and 2x2 on chip binning will enable this distinction to be made.  The longer integrations used resulted in a greater SNR for the 1x1 observations reducing the out of transit photometric scatter and reducing photometric noise by 36\%.   With a plate scale of 0.63$^{\prime\prime}$ the 1x1 observations are slightly over sampled for the typical seeing conditions producing a result similar to slightly defocussing the telescope but without some of the draw backs such as contamination from flux overlap.  The mid transit timing precision achievable depends on the RMS photometric noise,  the transit depth and the square root of the number of data points covering ingress/egress \citep[][Eq. 5]{Deeg2017TEETimes}.  As a  result, even though the unbinned observations were made at a 2.5 to 3 times lower cadence than the 2x2 on-chip binned observations, the significant reduction in RMS noise has a dominant effect reducing the transit mid-time uncertainty.

\begin{figure*}
 \includegraphics[width=1.0\textwidth]{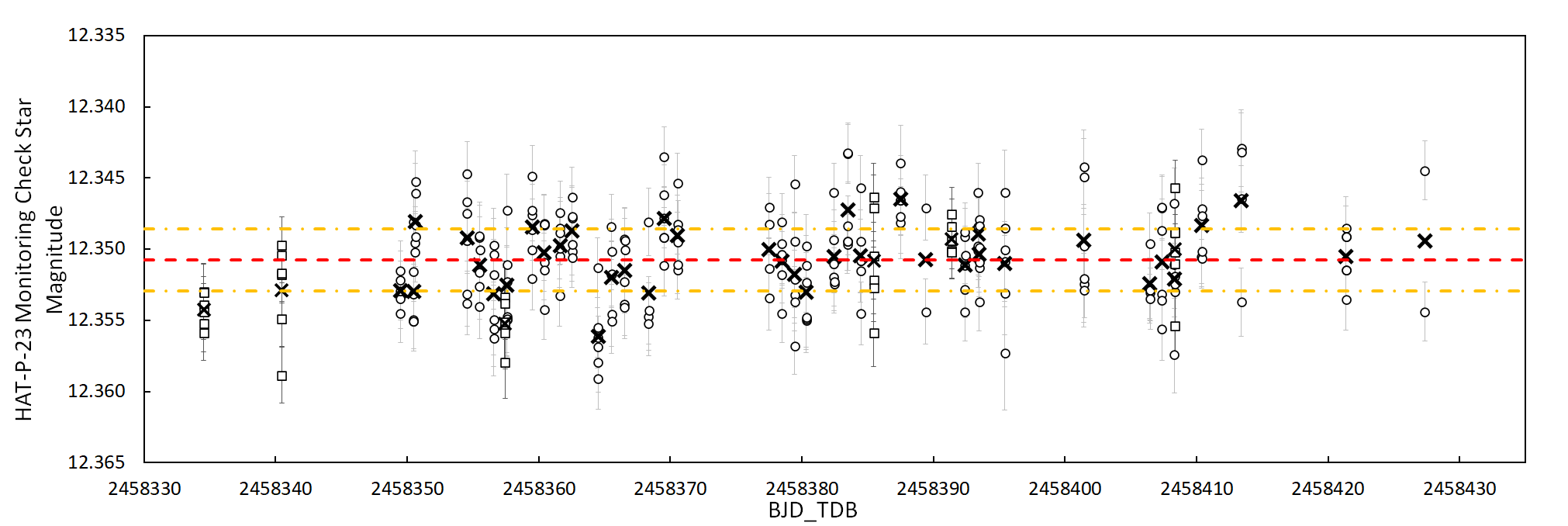}
 \caption{Nightly monitoring (open circles) and out of transit (open squares) photometry for the check star used to compare with HAT-P-23\,b. The crosses are the nightly mean values and the red dashed line is the mean value of the entire data set.  The dash-dot lines represent the $1\sigma$ variation of 0.002 mag over the 93 day period of observation.  }
 \label{HAT-P-23b_Check_Star_Photometry.png}
\end{figure*}

PIRATE was used to undertake monitoring of the HAT-P-23 field allowing us to analyse the  photometric precision obtained over a long duration.  The nightly observations were made with groups of 6 frames which exhibited a typical spread of 0.007 mag within a night.  The stability of the long term photometry is demonstrated with the standard error of the nightly mean magnitude which is 0.002 mag over 93 days, Figure \ref{HAT-P-23b_Check_Star_Photometry.png}. Of the 218 observations made, 95\% were obtained with airmass $\leq 1.5$ and no significant trend was found between measured magnitude of the check star and airmass. The magnitude measurements obtained from the out of transit data were found to have a -0.01 mag offset for the 2x2 on-chip binning compared to the un-binned observations which was corrected for prior to our analysis.  The monitoring observations conducted with PIRATE were clearly able to detect the variation in HAT-P-23 of 0.011 mag in measurements spanning 93 days.  Other suspected variable stars were detected in the field of HAT-P-23, the lowest amplitude being 0.005 mag with a 2.6 hour period.  These results demonstrate the capability of PIRATE to monitor both long and short period low amplitude variable stars.  

\begin{figure*}
 \includegraphics[width=1.0\textwidth]{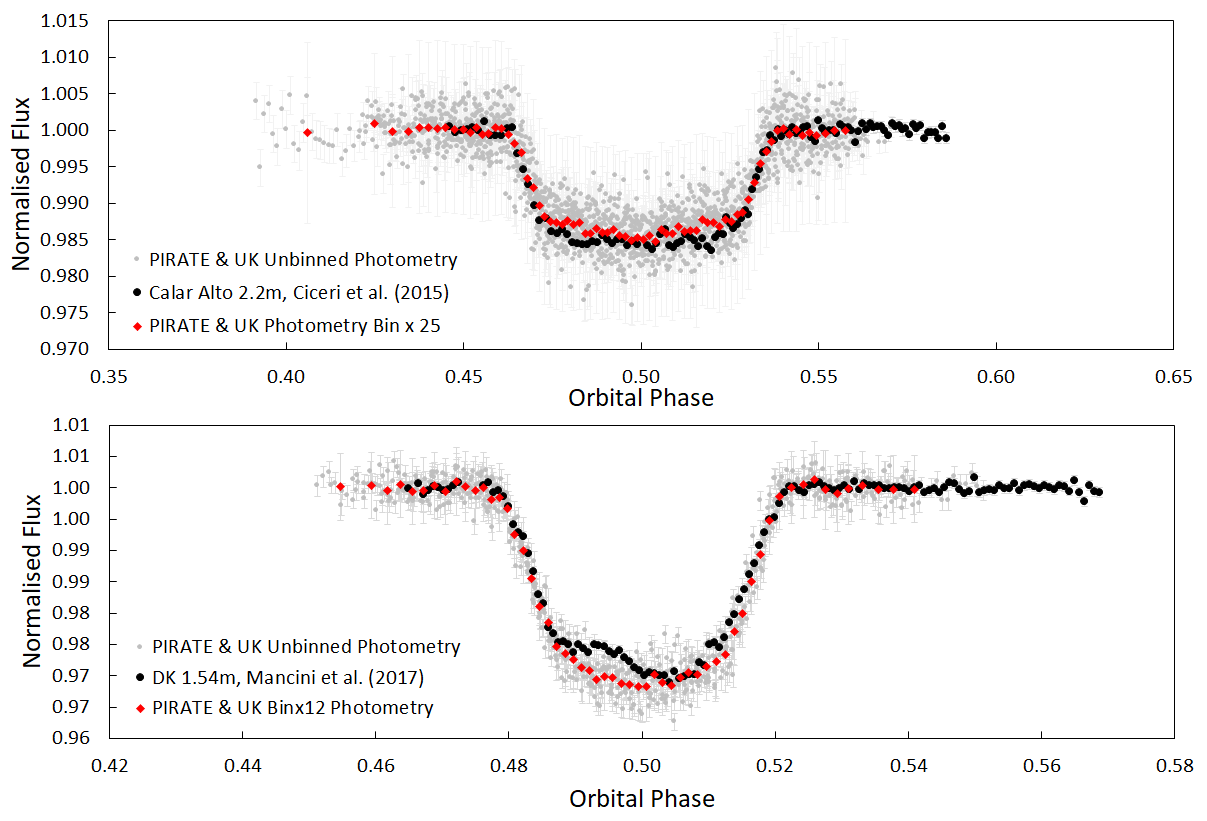}
 \caption{A comparison of the original and binned data for HAT-P-23\,b (top) and WASP-52\,b (bottom) with previously published results for medium class telescopes.  Clearly visible in the WASP-52\,b data from Mancini et al. (2017) is one of the spot crossing events observed.}
 \label{fig:Binned_Transit_Comp.png}
\end{figure*}

In Table \ref{tab:RMS scatter} we compare the photometric precision achieved with the small aperture telescopes used in this work with those from the literature using much larger aperture telescopes, and in Figure \ref{fig:Binned_Transit_Comp.png} we show our phase folded nightly and binned data for HAT-P-23b and WASP-52b  along with previously published observations made with the telescopes compared in Table \ref{tab:RMS scatter}.  The phase folded lightcurve for HAT-P-23b comprised 1821 individual measurements and a total observation time of 3337 minutes was binned by 25 data points.  By comparison \cite{Mancini2017OrbitalSystem} used the Calar Alto 2.2m telescope which has a primary mirror area approximately 24 times greater than the telescopes used in this study and obtained a single transit observation of 244 minutes duration.  Comparing telescope mirror area multiplied by the duration of observations we find that, in the absence of spot crossing events, the binned photometric results obtained with the small telescopes result in a lightcurve with the same or better photometric precision in half the mirror area $\times $ observation time of the single medium class telescope observation.  A similar result was found when the phase folded and binned WASP-52b transits were compared with those obtained using the Danish 1.54m telescope \cite{Ciceri2015}.  As the cost of a monolithic mirror telescope typically scales with aperture to the power \textbf{2.5 \citep{scopecostscaling}}, the method of combining the results obtained using multiple small telescope observations is an efficient approach to obtaining precise lightcurves for parameter determination of known transiting exoplanets with the added benefit of obtaining transit timing at multiple epochs.  

\section{Summary}\label{Summary}

We have presented 17 new transits for HAT-P-23b and 13 new transits of WASP-52b along with contemporaneous photometric monitoring for HAT-P-23, all made with 0.4m class telescopes.  For HAT-P-23b our combined radial velocity and new transit analysis results in an orbital eccentricity consistent with zero, in contrast the some previous studies \citep{Bakos2011HAT-P-20b-HAT-P-23b:Planets, Sada2016}, but consistent with the assumption of circular orbit by other authors \citep{Maciejewski2018Planet-starB, Ciceri2015}.  We find a radius for HAT-P-23b between previously published values but conclude that the planet should still be considered inflated, though the mechanism driving the inflation is not clear as the radius is smaller than that indicated by the empirical mass/incident flux relation for planetary radius derived by \cite{Weiss2013TheFlux}.  Our monitoring observations of HAT-P-23 have revealed a photometric modulation of  0.011 mag with a period of 7.015 days which we interpret as the rotation period of HAT-P-23 due to stellar surface spots.  

For WASP-52b our observations show no evidence for the large spot crossing events seen in previous works.  The detection of continuing brightness variations indicates that the spot latitude has migrated away from the transit chord.  Our analysis of the new and previously published transit times slightly prefers a quadratic ephemeris but only with a $\Delta \chi_{\nu}^2 =0.07$ and $\Delta$BIC = 1.53, \ref{tab:Ephemeris parameters}.  $\Delta$BIC values of less than 2 indicate no preference for one model over the other.  We have investigated the effects of removing different new and published transit timing data sets and find the quadratic model is preferred in all scenarios except where the very small uncertainties in the timing from \cite{Mancini2017OrbitalSystem} are taken at face value.  We conclude that apsidal precession and orbital decay are unlikely causes but that the Applegate effect cannot be ruled out.  However, considering the small statistical preference for the quadratic ephemeris and the high spot activity level of WASP-52, it seems the most plausible explanation for the slight preference of the quadratic ephemeris is increased scatter in transit timing measurements due to the effects of spot crossings.  Predicting the ephemerides forwards shows that transit  observations in the 2021 season will likely reveal if the quadratic ephemeris is real.

We have demonstrated that small aperture telescopes can be used to make valuable contributions to transiting exoplanet studies when deployed for long term monitoring campaigns.

\section*{Acknowledgements}
 This research was made possible through the OpenSTEM Labs, an initiative funded by HEFCE and by the Wolfson Foundation.  U.C.Kolb and C.A.Haswell are funded by STFC under consolidated grant ST/T000295/1.  This research has made use of the NASA Exoplanet Archive, which is operated by the California Institute of Technology, under contract with the National Aeronautics and Space Administration under the Exoplanet Exploration Program.

%%%%%%%%%%%%%%%%%%
% FIGURES
%%%%%%%%%%%%%%%%%%

%%%%%%%%%%%%%%%%%%
% TABLES
%%%%%%%%%%%%%%%%%%
\renewcommand{\arraystretch}{0.6}
\begin{landscape}
  \begin{table}  
    \caption{Table of observations made.  The line between observations 9 and 10 for HAT-P-23\,b indicates the point at which the upgrade to the PIRATE telescope was undertaken.  $N_{obs}$ is the number of observation per transit.  Dur.(min) is the total duration of the observations and Exp.(s) is the exposure time used.  RMS(ppt) is the out-of-transit residual RMS scatter from the transit fit in part per thousand.  CCD Mode indicates the on-chip binning used.  TSNR is the Transit Signal to Noise Ratio \citep{Ioannidis2016}.} 
    \label{tab:Observations}
                \begin{tabular}{ p{1.0cm}rlp{0.6cm}p{0.8cm}p{0.8cm}ccp{0.6cm}p{0.6cm}cl } 
        \hline\hline
       Number & Date & Telescope & Dur.
       (min) & Filter & Exp.(s) & $N_{obs}$ & Airmass & RMS
       (ppt) & CCD Mode & TSNR
       & Notes\\ 
       \hline\hline
       HAT-P-23b \\
       \hline\hline
        1&	05 October 2017	&POST	&205&	V&	150&	66&	1.27-2.41&	2.07&	2x2	&4.95 \\ 
        2&	20 June 2018&	PIRATE&	164&	Rc&	60&	158&	1.26-1.02&	2.04&	2x2&6.63&	Incomplete ingress coverage. \\ 
        3&	26 June 2018&	PIRATE&	213&	Rc&	60&	199&	1.14-1.02-1.11&	2.87&	2x2&4.71 \\	
        4&	30 June 2018&	PIRATE&	192&	Rc&	60&	178&	2.12-1.07&	3.29&	2x2&4.11 \\
        5&	30 June 2018&	POST &	180&	Rc&	120&	85&	1.58-1.21&	2.18&	2x2&6.19 \\
        6&	06 July 2018&	PIRATE&	216&	Rc&	60&	199	&1.33-1.02-1.04&	2.73&	1x1&4.95 \\
        7&	06 July 2018&	POST  &	264&	Rc&	120&	126&	1.55-1.21-1.27&	2.63&	2x2&5.13 \\
        8&	12 July 2018&	PIRATE&	210&	Rc&	60&	196	&1.05-1.02-1.25&	3.57&	2x2&3.78 \\
        9&	17 July 2018&	POST&	200&	Rc&	120&	94&	1.55-1.21&	2.19&	2x2&6.16 \\
        \hline
        10&	03 August 2018&	PIRATE&	196&	Rc&	180&	63&	1.41-1.02&	1.49&	1x1&9.05 \\	
        11&	09 August 2018&	PIRATE&	255&	Rc&	180&	67&	1.14-1.02-1.20&	0.85&	1x1&15.8&	Gap in data post egress. \\
        12&	26 August 2018&	PIRATE&	234&	Rc&	150&	92&	1.08-1.02-1.24&	1.85&	1x1&7.3 \\	
        13&	23 September 2018&	PIRATE&	195&	Rc&	180&	62&	1.06-1.02-1.17&	1.06&	1x1&12.71 \\	
        14&	23 September 2018&	POST&	181&	Rc&	180&	57&	1.21-1.49&	1.49&	2x2&9.08 \\	
        15&	29 September 2018&	PIRATE&	156&	Rc&	180&	51&	1.02-1.24&	1.55&	1x1&8.73&	Ingress only.  No transit mid time. \\
        16&	29 September 2018&	POST&	212&	Rc&	180&	53&	1.26-2.38&	2.32&	2x2&5.82&	Gap mid transit. \\
        17&	16 October 2018&	PIRATE&	229&	Rc&	180&	75&	1.03-2.12&	1.68&	1x1&8.02 \\	
        \hline\hline
       WASP-52b \\
        \hline\hline
        1&  10 November 2013 & POST & 148  & Rc & 90 & 93 & 1.44 - 1.40 & 2.06 & 3x3 &12.80 & Archive observation, spot crossing.\\
        2&	27 October 2017&	POST&	214&	Rc&	180&	66&	1.4-1.35-1.63&	1.60&	2x2&16.47\\	
        3&	01 August 2018&	PIRATE&	196&	Rc&	120&	92&	1.34-1.06&	1.64&	1x1&16.13	\\
        4&	08 August 2018&	PIRATE&	197&	Rc&	180&	55&	1.25-1.06-1.10&	1.80&	1x1&14.69&	Gap at egress.\\
        5&	29 August 2018&	PIRATE&	201&	Rc&	120&	95&	1.09-1.06-1.29&	1.62&	1x1&16.33\\	
        6&	19 September 2018&	PIRATE&	157&	Rc&	150&	60&	1.07-1.58&	2.02&	1x1&13.10\\	
        7&	03 October 2018&	PIRATE&	139&	Rc&	150&	52&	1.19-2.13&	2.07&	1x1&12.73\\	
        8&	05 October 2018&	POST&	183&	Rc&	120&	77&	1.78-1.35&	4.02&	2x2&6.57\\
        9&	10 October 2018&	PIRATE&	111&	Rc&	150&	43&	1.28-2.15&	- &	1x1& - &	No out of transit  coverage.\\
        10&	17 October 2018&	COAST&	133&	Rc&	240&	32&	1.21-2.17&	1.59&	1x1&16.65&	Ingress only.  No transit mid time.\\
        11&	17 October 2018&	PIRATE&	132&	B&	240&	33&	1.21-2.17&	2.40&	1x1&10.99&	Ingress only.  No transit mid time.\\
        12&	02 November 2018&	POST&	252&	Rc&	120&	113&	1.53-1.35-1.55&	2.15&	2x2&12.3\\	
        13&	07 December 2018&	POST&	110&	Rc&	240&	25&	1.36-1.55&	0.90&	1x1&29.33&	Ingress only.  No transit mid time.\\
        \hline\hline
        \end{tabular}
    \end{table}
\end{landscape}

%\begingroup
\renewcommand{\arraystretch}{1.15}
\begin{table*}
    \caption{Broadband photometry used in Exofastv2 SED fitting.} 
    \label{tab:SED photometry}
    \begin{center}
        \begin{tabular}{ lccl } 
       \hline\hline
       Band & HAT-P-23 & WASP-52&Catalogue\\ 
              \hline\hline
B(T)     & $13.362\pm\,{0.378}$& - &Tycho\\
V(T)     & $12.061\pm\,{0.195}$& - &Tycho\\
B      & $12.965\pm\,{0.036}$ & $13.116\pm\,{0.042}$ & APASS R10\\
V	   & $12.189\pm\,{0.047}$&$12.231\pm\,{0.062}$ & APASS R10\\
gSDSS&	$12.586\pm\,{0.043}$&$12.617\pm\,{0.025}$&APASS R10\\
rSDSS&	$11.969\pm\,{0.089}$&$11.901\pm\,{0.095}$&APASS R10\\
iSDSS&	$11.774\pm\,{0.143}$&$11.698\pm\,{0.047}$&APASS R10\\
J    & $11.103\pm\,{0.020}$&$10.588\pm\,{0.020}$&2MASS(via UCAC4)\\
H    & $10.846\pm\,{0.020}$&$10.186\pm\,{0.030}$&2MASS(via UCAC4)\\
K   &  $10.791\pm\,{0.020}$&$10.086\pm\,{0.030}$&2MASS(via UCAC4)\\
WISE1  & 10.748$^{+0.030}_{-0.023}$&$10.011^{+0.030}_{-0.023}$&AllWISE\\
WISE2 &  10.793$^{+0.030}_{-0.020}$&$10.094^{+0.030}_{-0.021}$&AllWISE\\
WISE3  & $10.816\pm\,{0.122}$&$9.983\pm\,{0.061}$&AllWISE\\
Gaia  &  12.174$^{+0.020}_{-0.001}$&$11.954^{+0.020}_{-0.001}$&Gaia DR2\\
GaiaBP&  12.518$^{+0.020}_{-0.002}$&$12.444^{+0.020}_{-0.004}$&Gaia DR2\\
GaiaRP & 11.679$^{+0.020}_{-0.002}$&$11.331^{+0.020}_{-0.003}$&Gaia DR2\\
           \hline\hline
        \end{tabular}
    \end{center}
\end{table*}
%\endgroup

 \begin{table*}
    \caption{HAT-P-23b transit times for complete new transits.  Transit numbers follow table 3.  All transit times used in the ephemeris calculations can be found in the supplementary material.} 
    \label{tab:HAT-P-23b transit times}
    \begin{center}
        \begin{tabular}{ cccccl} 
        \hline\hline
        Transit No.&Epoch&Tc (BJD$_{\textrm{TDB}}$)&Tc Uncertainty&Linear O-C&Telescope\\
        \hline\hline
1&2622&2458032.45353&0.00076&0.000075&POST\\
3&2839&2458295.65089&0.00057&0.001074&PIRATE\\
4&2843&2458300.50097&0.00075&-0.000392&PIRATE\\
5&2843&2458300.50190&0.00120&0.000538&POST\\
6&2848&2458306.56787&0.00071&0.002075&PIRATE\\
7&2848&2458306.56574&0.00070&-0.000055&POST\\
8&2853&2458312.62941&0.00078&-0.000817&PIRATE\\
9&2857&2458317.48136&0.00054&-0.000413&POST\\
10&2871&2458334.46187&0.00066&-0.000313&PIRATE\\
11&2876&2458340.5272&0.00045&0.000585&PIRATE\\
12&2890&2458357.50703&0.00060&0.000004&PIRATE\\
13&2913&2458385.40434&0.00068&0.000926&PIRATE\\
14&2913&2458385.40582&0.00075&0.002406&POST\\
16&2918&2458391.47024&0.00093&0.002393&POST\\
17&2932&2458408.44838&0.0006&0.000123&PIRATE\\
        \hline\hline
        \end{tabular}
    \end{center}
 \end{table*}

\begin{table*}
    \caption{WASP-52b transit times for complete new transits.  Results for both the linear and quadratic ephmerides are shown along with the difference between the two results.  Transit numbers follow table 3.  All transit times used in the ephemeris calculations can be found in the supplementary material.} 
    \label{tab:WASP-52b transit times}
    \begin{center}
        \begin{tabular}{ cccccccl} 
        \hline\hline
       Transit No.&Epoch&Tc (BJD$_{\textrm{TDB}}$) &Tc Uncertainty&Linear O-C&Quad O-C& $\Delta$ O-C&Telescope\\
        \hline\hline
2 & 1292 & 2458054.39890 & 0.00039 & -0.000194 & -0.000238 & 0.000044 & POST \\
3 & 1451 & 2458332.61424 & 0.00026 & -0.000049 & 0.000033 & -0.000082 & PIRATE \\
4 & 1455 & 2458339.61298 & 0.00031 & -0.000433 & -0.000347 & -0.000086 & PIRATE \\
5 & 1467 & 2458360.61083 & 0.00031 & 0.000044 & 0.000141 & -0.000097 & PIRATE \\
6 & 1479 & 2458381.60813 & 0.00036 & -0.000029 & 0.00008 & -0.00011 & PIRATE \\
7 & 1487 & 2458395.60619 & 0.00035 & -0.000218 & -0.0001 & -0.000118 & PIRATE \\
8 & 1488 & 2458397.35716 & 0.00054 & 0.000971 & 0.00109 & -0.000119 & POST \\
9 & 1491 & 2458402.60516 & 0.00076 & -0.000373 & -0.000251 & -0.000122 & PIRATE \\
12 & 1504 & 2458425.35296 & 0.00029 & 0.000273 & 0.000409 & -0.000136 & POST \\
        \hline\hline
        \end{tabular}
    \end{center}
 \end{table*}
 
\renewcommand{\arraystretch}{0.8}
\begin{table*}
    \caption{Median parameter values and 68\% confidence interval for WASP-52\,b and HAT-P-23\,b.} 
    \label{tab:Exofast outputs}
    \begin{center}
        \begin{tabular}{lccccccc}
        \hline\hline
        Parameter & Description  & WASP-52\,b & HAT-P-23\,b \\
        \hline\hline
$\rm{M}_*$\dotfill &Mass ($\rm{M}_{\odot}$)\dotfill &$0.844^{+0.042}_{-0.034}$&$1.063^{+0.063}_{-0.060}$\\
$\rm{R}_*$\dotfill &Radius ($\rm{R}_{\odot}$)\dotfill &$0.836\pm\,0.013$&$1.157^{+0.023}_{-0.022}$\\
$\rm{L}_*$\dotfill &Luminosity ($\rm{L}_{\odot}$)\dotfill &$0.399\pm\,0.013$&$1.460^{+0.087}_{-0.079}$\\
$\rho_*$\dotfill &Density (cgs)\dotfill &$2.04^{+0.14}_{-0.12}$&$0.967^{+0.075}_{-0.069}$\\
$\log{g}$\dotfill &Surface gravity (cgs)\dotfill &$4.52^{+0.024}_{-0.021}$&$4.338\pm\,0.028$\\
$\rm{T_{eff}}$\dotfill &Effective Temperature (K)\dotfill &$5017\pm\,41$&$5899^{+71}_{-68}$\\
$[{\rm Fe/H}]$\dotfill &Metallicity (dex)\dotfill &$0.13^{+0.11}_{-0.11}$&$0.150\pm\,0.015$\\
Age\dotfill &Age (Gyr)\dotfill &$8.5^{+3.7}_{-4.6}$&$5.5^{+3.0}_{-2.5}$\\
$A_{\rm{V}}$\dotfill &V-band extinction (mag)\dotfill &$0.060^{+0.034}_{-0.039}$&$0.132^{+0.079}_{-0.072}$\\
$d$\dotfill &Distance (pc)\dotfill &$173.1\pm\,1.6$&$360.0^{+6.0}_{-5.8}$\\
$\rm{R_P}$\dotfill &Planetary Radius ($\rm{R_J}$)\dotfill &$1.322\pm\,0.026$&$1.308^{+0.044}_{-0.043}$\\
$\rm{M_P}$\dotfill &Planetary Mass ($\rm{M_J}$)\dotfill &$0.446^{+0.025}_{-0.024}$&$1.92\pm\,0.13$\\
$a$\dotfill &Semi-major axis (AU)\dotfill &$0.02686^{+0.00044}_{-0.00037}$&$0.02273\pm\,0.00044$\\
$i$\dotfill &Inclination (Degrees)\dotfill &$84.90\pm\,0.24$&$83.60^{+0.67}_{-0.61}$\\
$e$\dotfill &Eccentricity \dotfill &$0.048^{+0.026}_{-0.027}$&$0.027^{+0.029}_{-0.019}$\\
$\omega_*$\dotfill &Argument of Periastron (Degrees)\dotfill &$85^{+33}_{-34}$&$178^{+93}_{-91}$\\
$T_{eq}$\dotfill &Equilibrium temperature (K)\dotfill &$1349^{+13}_{-14}$&$2029^{+29}_{-28}$\\
$\tau_{\rm circ}$\dotfill &Tidal circularisation timescale (Gyr)\dotfill &$0.00402^{+0.00052}_{-0.000546}$&$0.00435^{+0.00090}_{-0.00074}$\\
$K$\dotfill &RV semi-amplitude (m/s)\dotfill &$83.8^{+4.0}_{-3.9}$&$349^{+19}_{-20}$\\
$\rm{R_P/R_*}$\dotfill &Radius of planet in stellar radii \dotfill &$0.1625\pm\,0.0021$&$0.1162^{+0.0030}_{-0.0029}$\\
$\rm{a/R_*}$\dotfill &Semi-major axis in stellar radii \dotfill &$6.91^{+0.15}_{-0.13}$&$4.22^{+0.11}_{-0.10}$\\
$\delta$\dotfill &Transit depth (fraction)\dotfill &$0.02640\pm\,0.00069$&$0.01350^{+0.00069}_{-0.00067}$\\
$\tau$\dotfill &Ingress/egress transit duration (days)\dotfill &$0.01589^{+0.00073}_{-0.00071}$&$0.01243^{+0.00081}_{-0.00078}$\\
$\rm{T_{14}}$\dotfill &Total transit duration (days)\dotfill &$0.07795^{+0.00069}_{-0.00066}$&$0.09406^{+0.00089}_{-0.00088}$\\
$\rm{T_{FWHM}}$\dotfill &FWHM transit duration (days)\dotfill &$0.06206\pm\,0.00039$&$0.08161^{+0.00042}_{-0.00041}$\\
b\dotfill &Transit Impact parameter \dotfill &$0.590^{+0.019}_{-0.021}$&$0.472^{+0.039}_{-0.046}$\\
$\rho_P$\dotfill &Density (cgs)\dotfill &$0.239^{+0.020}_{-0.019}$&$1.06^{+0.13}_{-0.12}$\\
$\rm{logg_P}$\dotfill &Surface gravity \dotfill &$2.801\pm\,0.029$&$3.445\pm\,0.040$\\
$\Theta$\dotfill &Safronov Number \dotfill &$0.0214\pm\,0.011$&$0.0628^{+0.0041}_{-0.0040}$\\
${\langle F \rangle}$\dotfill &Incident Flux (cgs)\dotfill &$0.75\pm\,0.029$&$3.85^{+0.23}_{-0.21}$\\
$e\cos{\omega_*}$\dotfill & \dotfill &$0.002^{+0.023}_{-0.021}$&$-0.008^{+0.017}_{-0.030}$\\
$e\sin{\omega_*}$\dotfill & \dotfill &$0.042^{+0.027}_{-0.029}$&$-0.000^{+0.023}_{-0.024}$\\
$u_{1}$(B)\dotfill &linear limb-darkening coeff (B) \dotfill &$0.882^{+0.050}_{-0.051}$& -- \\
$u_{1}$(V)\dotfill &linear limb-darkening coeff (V) \dotfill & -- & $0.425\pm\,0.048$\\       
$u_{1}$(R)\dotfill &linear limb-darkening coeff (R) \dotfill & $0.532\pm\,0.018$ & $0.351\pm\,0.016$\\      
$u_{2}$(B)\dotfill &quadratic limb-darkening coeff (B) \dotfill &$-0.026\pm\,0.051$& -- \\        
$u_{2}$(V)\dotfill &quadratic limb-darkening coeff (V)\dotfill & -- & $0.252\pm\,0.049$\\
$u_{2}$(R)\dotfill &quadratic limb-darkening coeff (R)\dotfill & $0.180\pm\,0.017$ & $0.288\pm\,0.014$\\

        \hline\hline
        \end{tabular}
    \end{center}
\end{table*} 

\begin{table*}
\renewcommand{\arraystretch}{1.5}
    \caption{Parameter comparison for HAT-P-23\,b with TrEs-3\,b and WASP-135\,b which occupy the same location on a graph of planetary mass vs. radius vs.semi-major axis.  Values for HAT-P-23\,b from this work and TrEs-3\,b are taken from NASA Exoplanet Archive (accessed 2020-01-02). WASP-135\,b values are from \citep{Spake2016WASP-135b:Star}, except for $T_{eq}$ which was calculated from their published parameters using \citep[][Eq.5]{Southworth2010a}. See HAT-P-23\,b discussion.} 
    \label{tab:H23_T3_W135_Comp}
    \begin{center}
        \begin{tabular}{ cccc } 
        \hline\hline
        \textbf{Planet Parameter}& \textbf{HAT-P-23\,b} & \textbf{TrEs-3\,b} & \textbf{WASP-135\,b} \\ 
       \hline\hline
       $\rm{M_P~(M_J)}$ & $1.92\pm\,0.13 $ & $1.910^{+0.075}_{-0.080}$ & $1.90\pm\,0.08 $\\
       $\rm{R_P~(R_J)}$ & $1.308^{+0.044}_{-0.043} $ & $1.336^{+0.031}_{-0.037}$ & $1.30\pm\,0.09 $ \\
       a (AU) &  $0.02273\pm\,0.00044 $ & $0.02282^{+0.00023}_{-0.00040}$ & $0.0243\pm\,0.0005 $\\ 
       Age (Gyr) &  $5.5^{+3.0}_{-2.5} $ & $0.90^{+2.80}_{-0.80}$ & $0.60^{+1.40}_{-0.35} $\\
       $\rm{T_{eq}~(K)}$ &  $2029^{+29}_{-28} $ & $1638\pm\,22 $ & $1720\pm\,51 $\\
       $\langle \rm{F} \rangle~(10^9 erg~s^{-1} cm^{-2})$ &  $3.85^{+0.23}_{-0.21} $ & $1.650\pm\,0.079 $ & $1.98\pm\,0.24 $\\
        \hline\hline
        \end{tabular}
    \end{center}
\end{table*}

\begin{table*}
    \caption{Out of transit photometric RMS scatter for published observations with medium class telescopes compared with the phase folded and binned observations from this work.  ppm = parts per million.} 
    \label{tab:RMS scatter}
    \begin{center}
        \begin{tabular}{ lcccc } 
        \hline\hline
       \textbf{System} & \textbf{RMS Scatter (ppm)} & \textbf{Observation} & \textbf{Filter} & \textbf{Source} \\ 
       \hline\hline
        HAT-P-23\,b & 645 & Calar Alto 2.2m & Thuan-Gunn r & \cite{Ciceri2015} \\ 
        HAT-P-23\,b& 746 & Calar Alto 1.23m & Rc & \cite{Ciceri2015} \\ 
        HAT-P-23\,b & 415 & 0.4m (Binx25) & Rc & This work \\
        \hline
        WASP-52\,b & 582 & DK 1.54m & Bessell r & \cite{Mancini2017OrbitalSystem} \\ 
        WASP-52\,b & 768 & Calar Alto 1.23m & Rc & \cite{Mancini2017OrbitalSystem} \\ 
        WASP-52\,b & 616 & 0.4m (Binx12) & Rc & This work \\
        \hline\hline
        \end{tabular}
    \end{center}
\end{table*}

\begin{table*}
     \caption{Ephemeris values for WASP-52\,b and HAT-P-23\,b} 
    \label{tab:Ephemeris parameters}
    %\begin{centre}
        \hspace*{-4cm}\begin{tabular}{llccccc}
        \hline\hline
        Planet & Ephemeris & Epoch & Period & Quadratic Term & $\chi_{\nu}^2$ & BIC \\
        \hline\hline
WASP-52\,b & Linear & $2455793.681914(141)$ & $1.749781099(126)$ & -- & 1.58 & 59.41\\ 
WASP-52\,b & Quadratic & $2455793.680977(141)$ & $1.749783241(177)$ & $-2.14\pm\,0.24\times10^{-9}$ & 1.51 & 57.88\\
\hline
HAT-P-23b & Linear & $2454852.265165(120)$ & $1.212886457(54)$ & -- & 2.08 & 33.01\\
         \hline\hline
        \end{tabular}
    %\end{centre}
 \end{table*}

%\section*{References}
\clearpage
\bibliography{references2.bib}

\end{document}